\documentclass[times,authoryear]{elsarticle}

\usepackage{jasr}
\usepackage{framed,multirow}
\usepackage{amssymb,latexsym,amsmath,amsfonts,mathtools,bm}
\usepackage{array}
\usepackage{booktabs}
\usepackage{graphicx}
\usepackage{algorithm}
\usepackage{algpseudocode}
\usepackage{url}
\usepackage{xcolor}
\definecolor{newcolor}{rgb}{.8,.349,.1}
\usepackage[citebordercolor=white]{hyperref}
\usepackage[switch]{lineno}
\usepackage{svg}
\usepackage{verbatim}
\usepackage{xfrac}
\usepackage{subcaption}
\journal{Advances in Space Research}

\numberwithin{equation}{section}
\newcommand{\transpose}{\mathsf{T}}
\newcommand{\diag}{\mathrm{diag}}
\newcommand{\norm}[1]{\left\lVert #1 \right\rVert}
\newcommand{\abs}[1]{\left\lvert #1 \right\rvert}

\begin{document}

\begin{frontmatter}

\title{Adaptive Orbit-Aided Vector Tracking for Orbit Determination of LEO Satellites}

\author[1]{Zihong Zhou}
\author[1]{Bing Xu\corref{cor1}}
\cortext[cor1]{Corresponding author}
\ead{pbing.xu@polyu.edu.hk}
\affiliation[1]{organization={Department of Aeronautical and Aviation Engineering, The Hong Kong Polytechnic University},
                addressline={Hung Hom, Kowloon},
                city={Hong Kong SAR},
                country={China}}

\begin{abstract}

Reliable orbit determination for CubeSats is essential for emerging low-Earth-orbit positioning, navigation, and timing (LEO-PNT) applications. However, onboard orbit determination remains constrained by single-frequency GNSS code-based receivers, intermittent signal reception, and highly dynamic orbital conditions. This paper proposes an adaptive orbit-aided vector tracking (OAVT) algorithm for LEO-PNT CubeSat orbit determination under these conditions. The method employs a GNSS vector tracking loop (VTL) as the primary navigation estimator and loosely couples it with a low-fidelity orbital dynamics model (ODM), which provides auxiliary position and velocity pseudo-measurements. The contribution of the ODM is regulated through adaptive covariance assignment based on GNSS stress and the residual consistency between the ODM and GNSS solutions. Under nominal conditions, OAVT preserves the accuracy advantage of VTL over the conventional scalar tracking loop. The 3D position root-mean-square error (RMSE) is reduced from 2.631\,m for STL to 1.081\,m, while the 3D velocity RMSE decreases from 0.3985\,m/s to 0.1023\,m/s. Under degraded conditions, OAVT provides substantially greater robustness than both STL and standalone VTL. The 3D position RMSE remains bounded at 8.27\,m during a complete signal outage, 10.17\,m during gradual signal loss and reacquisition, and 26.65\,m under variable visibility, whereas STL fails to produce valid solutions and standalone VTL loses lock. These results demonstrate that OAVT substantially improves solution robustness and state continuity with only a moderate increase in computational burden, making it suitable for CubeSat orbit determination and future LEO-PNT missions.

\end{abstract}


\end{frontmatter}

%\linenumbers

\section{Introduction}
\label{sec:intro}

Low-Earth orbit (LEO) is increasingly recognised as a critical domain for future positioning, navigation, and timing (PNT) augmentation architectures. Compared to conventional medium Earth orbit (MEO) global navigation satellite systems (GNSSs), LEO-based systems offer stronger signal power, more rapid geometric diversity, and the prospect of enhanced resilience and faster convergence in demanding PNT applications \citep{morton2021,prol2022,he2026enhancing}. In this context, CubeSats and related low-cost small satellite platforms have emerged as particularly attractive candidates for the construction of future LEO-PNT constellations, owing to their potential for reduced mission costs, faster deployment, and scalable constellation growth. Nevertheless, these platforms inherently impose stringent constraints on size, weight, power consumption, onboard computational resources, and unit cost \citep{allahvirdi2022pod,ge2022}.

The quality of the PNT services provided, whether augmentation-based or standalone, ultimately depends on the precise and timely knowledge of the LEO satellites' orbits and clocks \citep{kunzi2025gnss}. For large-class satellites equipped with sophisticated instrumentation, a well-developed spectrum of precise orbit determination (POD) methodologies exists, including orbital dynamic model (ODM) propagation, satellite laser ranging (SLR), Doppler orbitography and radiopositioning integrated by satellite (DORIS), and precise point positioning (PPP) supported by precise GNSS products \citep{selvan2021review2}. These approaches are capable of achieving metre-, decimetre-, and even centimetre-level POD accuracy, given sufficient processing capability \citep{montenbruck2005reduced,choi2011nrtpod,selvan2023podreview}. For CubeSats and other miniaturised platforms, however, the reliance on commercial off-the-shelf (COTS) components typically restricts onboard processing to single-frequency, code-based kinematic POD derived from GNSS broadcast ephemerides \citep{prol2022,allahvirdi2022pod,selvan2023podreview}.

Within this line of research, early work on kinematic POD studied real-time orbit determination from GNSS observables with lightweight Kalman filtering \citep{gomes2007rtod,chiaradia2003sfrt,hwang2005sf}. Single-frequency GNSS strategies with ionospheric mitigation were then demonstrated to support metre-level orbit recovery for LEO satellites under favourable conditions \citep{chen2017sf,sun2017real,tseng2024satellite}. More recent small-satellite studies have further improved the availability and robustness of single-frequency kinematic solutions under realistic onboard constraints, such as clock jumps, attitude-dependent measurement quality, and interrupted observation continuity \citep{fu2021sfkod}. To date, however, the accuracy of broadcast ephemeris-based single-frequency kinematic POD for CubeSats remains limited to the metre level \citep{omar2018singlefreq,allahvirdi2022thesis,gangestad2013operations,selvan2021review2}, a performance regime widely regarded as insufficient for LEO-PNT applications \citep{teunissen2017handbookgnss}.

Both the LEO signal environment and the practical limitations of CubeSat-grade hardware make spaceborne GNSS POD distinctively challenging. Due to the high relative line-of-sight (LOS) dynamics between LEO and GNSS satellites, the GNSS signals received by spaceborne receivers are subject to Doppler shifts of up to $\pm$60,000 Hz and rapidly varying carrier-to-noise ratio ($C/N_0$) ranging from 10 to 55 dB-Hz \citep{samba2025iac,montenbruck2008gras,allahvirdi2022pod}. The resulting measurement inconsistency and occasional signal outages impose considerable difficulty in maintaining stable tracking within the conventional scalar tracking loop (STL) receivers, where the receiver's own dynamics remain uncompensated during tracking, leaving the loop directly susceptible to high Doppler and unstable signal strength \citep{xu2019vt,lashley2007stlvtlcom}. In addition, CubeSats typically rely on temperature-compensated or oven-controlled crystal oscillators (TCXOs/OCXOs) for onboard timekeeping; the poor stability of these oscillators necessitates frequent estimation and correction of clock bias and drift, which becomes especially critical for navigation message broadcast \citep{wang2022clock}.

In the context described above, the vector tracking loop (VTL) receiver architecture presents a compelling alternative. By aggregating power levels of all channels and employing a Kalman filter (KF) for joint state estimation, VTL allows signals with higher $C/N_0$ to exert greater influence on the final solution than weaker signals \citep{petovello2009}. Moreover, by feeding back the estimated navigation states to predict LOS parameters for the subsequent epoch, VTL enables stronger channels to assist in sustaining tracking of weaker or temporarily degraded signals. Owing to these characteristics, VTL exhibits improved robustness in weak-signal and high-dynamic scenarios relative to the conventional STL \citep{lashley2009,xu2019vt}. 

The application of VTL to the LEO spaceborne scenario has been preliminarily investigated and demonstrated to be feasible by \citet{liao2024leo}. Nevertheless, the existing literature lacks a systematic performance comparison against conventional receivers, and several issues critical to the LEO-PNT context remain unaddressed. First, due to various factors including satellite orbit (e.g., polar orbits), satellite attitude, antenna performance, antenna stabilisation problems, and specific channel assignment, GNSS satellite visibility can fluctuate markedly, resulting in degraded tracking continuity or, in severe cases, intermittent signal outages \citep{wang2020dutycycle,reid2016polar}. \citet{zhao2020tianping} reported that fewer than four BeiDou Navigation Satellite System-3 (BDS-3) satellites were visible in an SSO for roughly 37\% of epochs, with none visible at all for 3.3\% of the time. \citet{wang2020dutycycle} further reported that, even for a dual-frequency LEO CubeSat, the percentage of valid single-point-positioning (SPP) solutions could be as low as 82.9\%. Second, the information sharing within VTL renders it susceptible to cross-channel contamination, whereby erroneous measurements acquired during periods of poor visibility can corrupt the aggregate navigation solution and subsequently compromise tracking performance across all channels if not adequately managed \citep{petovello2009}. Third, duty-cycling among onboard instruments is a common operational practice necessitated by the constrained power budgets of CubeSats \citep{wang2020dutycycle}. For conventional scalar tracking loop receivers, off-duty intervals mean that no updated orbit or clock information can be produced. Whilst a VTL receiver is better suited than STL in this regard, as the navigation filter can continue to propagate state estimates in the absence of GNSS measurements and achieve rapid reacquisition upon resumption of tracking, the resulting propagation errors nonetheless accumulate rapidly with increasing outage duration \citep{xu2019vt}. Terrestrial users of LEO-PNT services, however, continue to demand accurate and timely satellite position information throughout such intervals \citep{morton2021,selvan2023podreview}. Taken together, these limitations indicate the need for an orbit-determination framework that fully exploits VTL in the spaceborne high-dynamic regime, while still supporting frequent estimation of onboard clock parameters from GNSS measurements. At the same time, such a framework must remain compatible with single-frequency, code-based processing driven by broadcast ephemerides, preserve acceptable PNT accuracy during receiver off-duty intervals, degraded-observation periods, and corrupted-measurement events, and do so with sufficiently low computational burden for practical CubeSat implementation.

To address the research gaps, and inspired by the methods of multi-sensor GNSS fusion with other systems \citep{wang2024mpkf,cui2024context,mohamed1999adaptive}, this paper proposes the adaptive orbit-aided vector tracking algorithm, referred to hereafter as OAVT. OAVT is implemented as a software-defined receiver (SDR) for LEO-PNT CubeSat orbit determination, in which the VTL remains the primary estimator so that onboard clock bias and drift can continue to be updated frequently from GNSS measurements, while a low-fidelity ODM is loosely coupled as an auxiliary source to provide short- to mid-term position-velocity constraints at low computational cost (hence the name "orbit-aided"). The navigation states are first produced independently by the VTL and the ODM, after which the ODM-propagated position and velocity are injected into a KF estimator as pseudo-measurements for solution fusion, with the associated covariance adjusted adaptively based on a GNSS stress scale computed from the root-mean-square (RMS) observable innovations and a geometric dilution of precision (GDOP) penalty, together with the residual disagreement between the ODM-propagated and GNSS-estimated states \citep{li2022adaptivepod,cheng2024adaptivecarrier}; residual gating is further applied to screen out inconsistent ODM updates prior to the final KF solution fusion. In this way, the ODM imposes a physically informed constraint on the POD solution under changing GNSS observation conditions and receiver off-duty intervals, so that error growth can be suppressed to maintain adequate positioning quality for terrestrial users whilst simultaneously facilitating rapid GNSS reacquisition upon measurement resumption. Because the ODM is intended only to support the VTL over short- to mid-duration outages, it is deliberately kept low fidelity to minimise computational cost. The loose-coupling architecture and adaptive covariance scheme collectively ensure that, should the VTL fail or yield degraded solutions, the ODM can independently provide uncontaminated state estimates and restore the VTL within a single propagation step, whilst the residual gating mechanism reciprocally safeguards solution integrity against anomalous ODM outputs.

The main contributions of this work are summarised as follows:
\begin{enumerate}
    \item A spaceborne orbit-determination framework is developed that fully exploits the VTL architecture for the high-dynamic LEO-PNT setting under the practically relevant CubeSat regime of single-frequency, code-based GNSS processing using broadcast ephemerides.
    \item A computationally efficient orbit-aiding strategy is established by integrating a low-fidelity ODM into the GNSS receiver, enabling sustained PNT performance during receiver off-duty intervals, degraded-observation periods, and corrupted-measurement events, while also supporting rapid tracking recovery after measurement resumption.
    \item An adaptive integration framework is formulated in which the VTL serves as the primary estimator for frequent GNSS-based clock parameter updates, while a loosely coupled ODM provides auxiliary position-velocity constraints through pseudo-measurement fusion, with the associated covariance assigned adaptively according to GNSS solution quality.
\end{enumerate}

The remainder of this paper is organised as follows. Section \autoref{sec:methodology} details the algorithm methodology, including the overall architecture, the GNSS error-state navigation filter, the low-fidelity ODM, and the adaptive OAVT integration and update scheme. Section \autoref{sec:results} presents the performance results under nominal and degraded conditions using simulated spaceborne GNSS signals. Section \autoref{sec:discussion} discusses computational complexity and algorithm parameter tuning. Section \autoref{sec:conclusion} concludes the paper.

\textbf{Notation.} Throughout this paper, a hat accent $\hat{(\cdot)}$ denotes an estimated quantity, while a breve accent $\breve{(\cdot)}$ denotes a measured quantity. The superscripts $(\cdot)^-$ and $(\cdot)^+$ denote, respectively, the \textit{a priori} and \textit{a posteriori} states of a given process. The operator $\|\cdot\|$ denotes the Euclidean norm of a vector. The operators $\odot$ and $\oslash$ denote element-wise multiplication and division, respectively. The operator $\mathrm{clip}(x, a, b) \triangleq \max\!\left(a, \min(b, x)\right)$ denotes scalar clamping to the interval $[a, b]$, applied element-wise when the argument is a vector.

\section{Methodology}
\label{sec:methodology}

This section presents the framework and implementation details of adaptive OAVT. The GNSS receiver employs a vector delay lock loop (VDLL) architecture and is implemented on the platform of the GSRx Software GNSS Receiver developed by the Finnish Geospatial Research Institute (FGI) \citep{soderholm2022fgigsrx} and the open-source MATLAB code for a vector tracking-based GNSS SDR developed by \citet{xu2019vt}. The architecture overview and coupling scheme of the OAVT algorithm are provided in \autoref{subsec:algarch}; the GNSS extended Kalman filter (EKF) design and measurement estimation are detailed in \autoref{subsec:ekf}; the ODM principle is explicated in \autoref{subsec:odm}; the adaptive pseudo-measurement integration mechanism is elaborated in \autoref{subsec:couplingfilter}.

\subsection{Algorithm architecture}
\label{subsec:algarch}

\begin{figure}[htbp]
    \centering
    \includegraphics[width=0.8\linewidth]{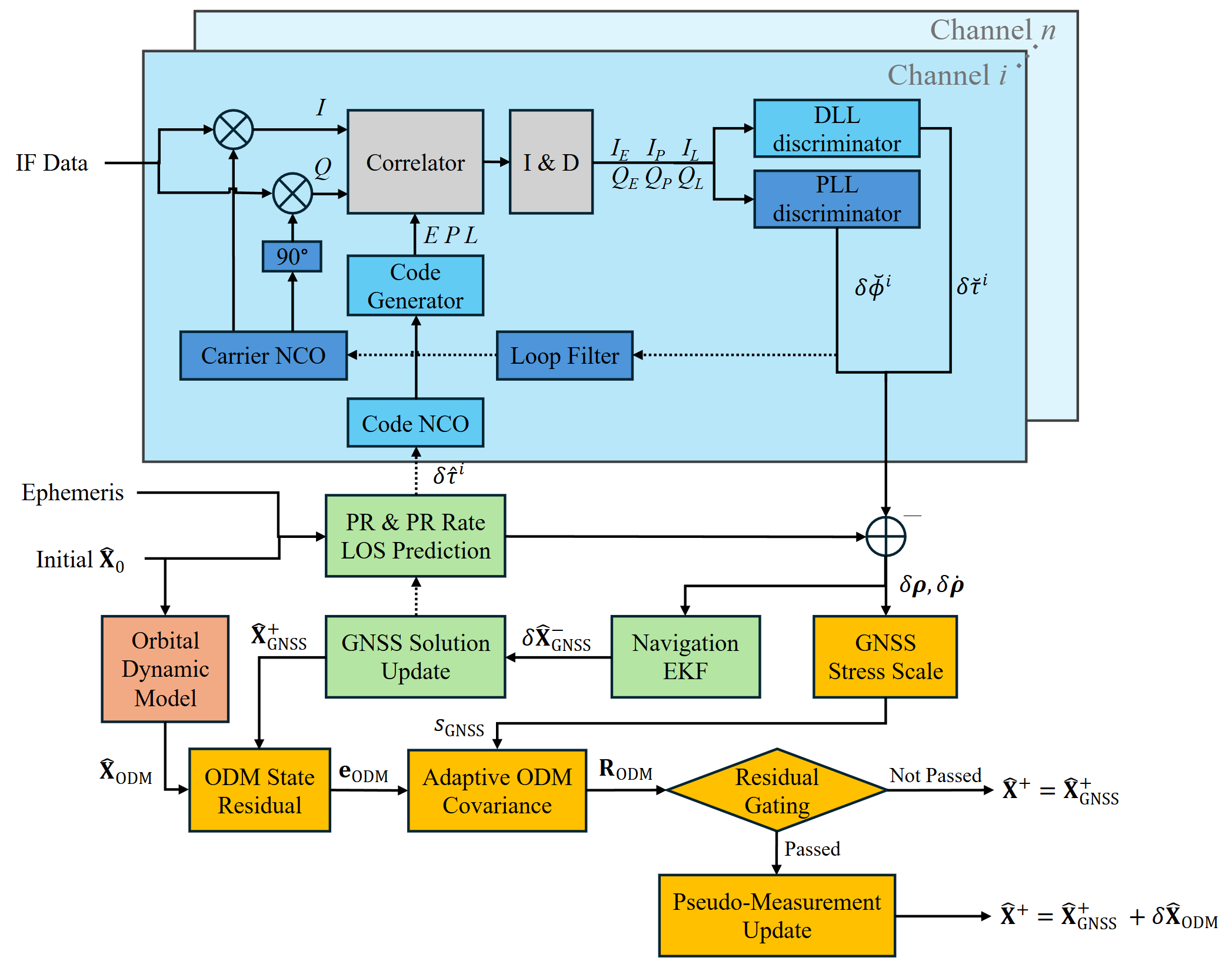}
    \caption{The algorithm architecture of the OAVT.}
    \label{fig:schem}
\end{figure}

The overall architecture of the OAVT is illustrated in \autoref{fig:schem}. The architecture is organised around two interacting loops operating at different time scales: a high-rate vector tracking navigation loop executed at every predetection integration (PDI) epoch, and a lower-rate orbit-aiding loop executed at scheduled ODM update epochs. The high-rate VTL remains the primary source of navigation state and clock estimates, whereas the low-rate ODM contributes dynamically plausible position and velocity information when the GNSS solution becomes stressed. This multi-rate arrangement preserves the closed-loop nature of VTL while introducing a computationally efficient orbital constraint suitable for CubeSat onboard processors. 

At the channel level, the incoming intermediate frequency (IF) samples are mixed into the in-phase and quadrature branches, correlated with the locally generated early, prompt, and late code replicas, and accumulated to form the in-phase and quadrature early, prompt, and late correlator values. These quantities are then passed to the delay-locked loop (DLL) and phase-locked loop (PLL) discriminators, which produce residual code-delay error $\delta \breve{\tau}$ and carrier-phase error $\delta \breve{\phi}$. Through the loop filter, these residuals update the local numerically controlled oscillator (NCO) commands, thereby maintaining channel lock.

The discriminator outputs are then converted into the measurement residuals $\delta \rho$ and $\delta \dot{\rho}$, which drive an error-state navigation EKF. The filter produces the GNSS state correction $\delta \hat{\mathbf{X}}_{\mathrm{GNSS}}$ and the updated GNSS solution $\hat{\mathbf{X}}^{+}_{\mathrm{GNSS}}$, which is subsequently used to compute the predicted LOS pseudorange and pseudorange-rate, and finally, the predicted code error $\delta \hat{\tau}$ for aiding the tracking in the next epoch. To initialise the loop, the decoded GNSS broadcast ephemerides and the initial spacecraft state $\hat{\mathbf{X}}_{0}$ need to be provided by a conventional STL.

In parallel to the high-rate GNSS loop, the initial state $\hat{\mathbf{X}}_{0}$ is also supplied to the ODM branch, which propagates the spacecraft trajectory and returns the orbital projection $\hat{\mathbf{X}}_{\mathrm{ODM}}$. This propagated dynamic solution enters the adaptive covariance and residual-gating stages before reaching the solution fusion filter. When the GNSS innovations increase or the measurement geometry deteriorates, the covariance assigned to the ODM pseudo-measurement is reduced so that the correction carries greater authority.

\begin{figure}[htbp]
    \centering
    \includegraphics[width=0.7\linewidth]{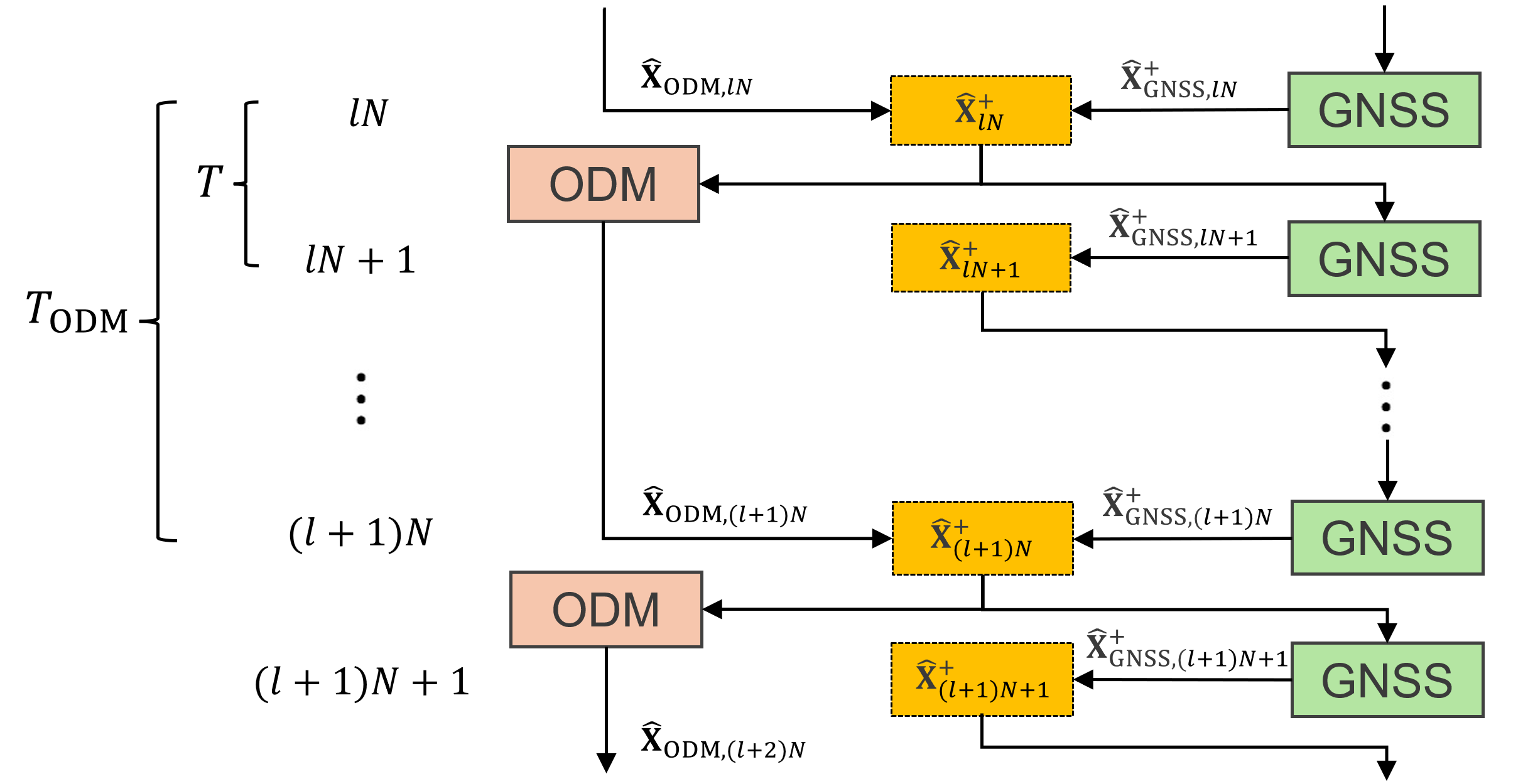}
    \caption{The coupling schedule between the GNSS and ODM loops.}
    \label{fig:coupling}
\end{figure}

The ODM updates at its scheduled epochs, and the resulting coupling schedule between the GNSS and ODM loops is illustrated in \autoref{fig:coupling}. The GNSS loop advances at every epoch $k$ with an update interval of $T$, whereas the ODM is invoked only once every $N_{\mathrm{ODM}}$ epochs (abbreviated $N$ hereafter), giving an update period $T_{\mathrm{ODM}} = N T$. Indexing successive ODM updates by the counter $l = 1, 2, 3, \dots$, the $l$-th update is initiated at GNSS epoch $k = l N$, where the orbit is propagated forward for $T_{\mathrm{ODM}}$ to yield the predicted state $\hat{\mathbf{X}}_{\mathrm{ODM},\, l+1}$ at the next ODM scheduled epoch $k = (l+1) N$; the two time scales therefore coincide only at these fusion epochs. Between scheduled updates, the receiver runs on the GNSS-only solution, $\hat{\mathbf{X}}^{+}_k = \hat{\mathbf{X}}^{+}_{\mathrm{GNSS},k}$, and only at $k = l N$ is the ODM prediction admitted as a pseudo-measurement for fusion: if the residual gate is passed, the solution is fused as $\hat{\mathbf{X}}^{+}_{lN} = \hat{\mathbf{X}}^{+}_{\mathrm{GNSS},lN} + \delta \hat{\mathbf{X}}_{\mathrm{ODM},lN}$; otherwise, the architecture retains the GNSS-only solution, $\hat{\mathbf{X}}^{+}_{lN} = \hat{\mathbf{X}}^{+}_{\mathrm{GNSS},lN}$. Through this coupling schedule, the orbital constraint is imposed at a reduced rate without breaking the closed loop of the high-rate tracking.

\subsection{GNSS EKF design and measurement estimation}
\label{subsec:ekf}

\subsubsection{State vector and EKF propagation}
The GNSS navigation filter is implemented in error-state form. As the spaceborne scenario is considered highly dynamic, the navigation state is an 11-state vector expressed in the Earth-Centred Earth-Fixed (ECEF) frame,

\begin{equation}
    \delta \hat{\mathbf{X}}^-_k
    =
    \begin{bmatrix}
        \delta \hat{x},\ \delta \hat{y},\ \delta \hat{z},\ \delta\hat{v}_x,\ \delta \hat{v}_y,\ \delta \hat{v}_z,\ \delta \hat{a}_x,\ \delta \hat{a}_y,\ \delta \hat{a}_z,\ c\delta \hat{t}_{r},\ c\delta \hat{\dot{t}}_{r}
    \end{bmatrix}_k^{\transpose},
    \label{eq:state11}
\end{equation}
where $k$ denotes the current epoch, $\delta \hat{\mathbf{r}} = [\delta x,\delta y,\delta z]^{\transpose}$ is the position error vector; $\delta \hat{\mathbf{v}} = [\delta \hat{v}_x, \delta \hat{v}_y,\delta \hat{v}_z]^{\transpose}$ is the velocity error vector; $\delta \hat{\mathbf{a}} = [\delta \hat{a}_x, \delta \hat{a}_y,\delta \hat{a}_z]^{\transpose}$ is the acceleration error vector; $c$ denotes the speed of light; and $c\delta \hat{t}_{r,k}$ and $c\delta \hat{\dot{t}}_{r,k}$ are the receiver clock bias and drift, respectively.

Over one update interval $T$, the kinematic states follow a constant-acceleration model and the clock states follow a constant-drift model. The resulting discrete-time transition matrix is
\begin{equation}
    \mathbf{F}
    =
    \begin{bmatrix}
        \mathbf{I}_3 & T\mathbf{I}_3 & \tfrac{T^2}{2}\mathbf{I}_3 & \mathbf{0}_{3\times1} & \mathbf{0}_{3\times1} \\
        \mathbf{0}_{3\times3} & \mathbf{I}_3 & T\mathbf{I}_3 & \mathbf{0}_{3\times1} & \mathbf{0}_{3\times1} \\
        \mathbf{0}_{3\times3} & \mathbf{0}_{3\times3} & \mathbf{I}_3 & \mathbf{0}_{3\times1} & \mathbf{0}_{3\times1} \\
        \mathbf{0}_{1\times3} & \mathbf{0}_{1\times3} & \mathbf{0}_{1\times3} & 1 & T \\
        \mathbf{0}_{1\times3} & \mathbf{0}_{1\times3} & \mathbf{0}_{1\times3} & 0 & 1
    \end{bmatrix}.
    \label{eq:F11}
\end{equation}
The predicted absolute state is therefore
\begin{equation}
    \delta\hat{\mathbf{X}}^-_k = \mathbf{F}\delta\hat{\mathbf{X}}^+_{k-1}.
    \label{eq:absolute_prediction}
\end{equation}

In the implementation, the process noise covariance is kept diagonal for computational efficiency,
\begin{equation}
    \mathbf{Q} = \text{diag}\!\left(
    q_{\delta x},\ q_{\delta y},\ q_{\delta z},\
    q_{\delta\dot{x}},\ q_{\delta\dot{y}},\ q_{\delta\dot{z}},\
    q_{\delta\ddot{x}},\ q_{\delta\ddot{y}},\ q_{\delta\ddot{z}},\
    q_{c\delta t_r},\ q_{c\delta\dot{t}_r}
    \right),
    \label{eq:Qdiag}
\end{equation}
where the terms are tuned empirically according to the expected receiver dynamics and clock stability \citep{mu2021vtl,xu2019vt}.

The error-state covariance is propagated as
\begin{equation}
    \mathbf{P}^-_k = \mathbf{F}\mathbf{P}^+_{k-1}\mathbf{F}^{\transpose} + \mathbf{Q}.
    \label{eq:cov_prediction}
\end{equation}

\subsubsection{Pseudorange and pseudorange-rate error formation}
\label{subsec:prerror}

Correspondingly, the pseudorange observable enters the error-state EKF in the form of a measurement residual, defined as the difference between the observed and predicted pseudorange. For the $i$-th GNSS satellite at epoch $k$, the pseudorange residual is
\begin{equation}
    \delta \rho^i_{k} = \breve{\rho}^i_{k} - \hat{\rho}^i_{k},
    \label{eq:pr_error_raw}
\end{equation}
The predicted pseudorange $\hat{\rho}^{i}_{k}$ is constructed from the observation model \citep{enge1994global}

\begin{equation}
    \hat{\rho}^i_{k}
    =
     \norm{
        \mathbf{r}^{i}_{s,k} - \hat{\mathbf{r}}^{-}_{k}
    }
    +
    c\delta \hat{t}^{-}_{r,k}
    +
    c\delta t^i_{s,k}
    -
    c\delta t^{i}_{\mathrm{grp},k}
    -
    I^i_{k},
    \label{eq:pred_pr}
\end{equation}
where $\mathbf{r}^{i}_{s,k}$, $c\delta t^i_{s,k}$, and $c\delta t^{i}_{\mathrm{grp},k}$ denote the broadcast GNSS satellite position, clock bias, and group delay, respectively; $I^i_{k}$ is the ionospheric delay correction. Since the receiver operates above the Earth's troposphere, the tropospheric delay is absent and its correction term is omitted. In the VDLL receiver, the measurement residual can be directly approximated from the code phase discriminator output $\delta \breve{\tau}^i_k$ in chips as
\begin{equation}
    \delta \rho_{i,k}
    =
    \frac{c}{f_{\mathrm{code}}} \delta \breve{\tau}^i_k.
    \label{eq:pr_error_dll}
\end{equation}
where $f_{\mathrm{code}}$ is the nominal code chipping rate.

The pseudorange-rate residual is handled analogously. For the $i$-th GNSS satellite at epoch $k$, the pseudorange-rate residual is defined as
\begin{equation}
    \delta \dot{\rho}^i_{k}
    =
    \breve{\dot{\rho}}^i_{k}
    -
    \hat{\dot{\rho}}^i_{k}
    -
    c\delta\hat{\dot{t}}^-_{r,k}
    +
    c\delta \dot{t}^{i}_{s,k}.
    \label{eq:prr_error}
\end{equation}
where $c\delta \dot{t}^{i}_{s,k}$ is the broadcast GNSS satellite clock drift. The measured pseudorange-rate is derived from the carrier Doppler estimate as
\begin{equation}
    \breve{\dot{\rho}}^i_{k}
    =
    \left(
        \breve{f}^{i}_{\mathrm{carr},k}
        -
        f_{\mathrm{IF}}
    \right)
    \frac{c}{f_{\mathrm{carr}}},
    \label{eq:meas_prr}
\end{equation}
where $\breve{f}^{i}_{\mathrm{carr},k}$ is the tracked carrier frequency; $f_{\mathrm{IF}}$ is the receiver IF reference after centre-frequency adjustment; and $f_{\mathrm{carr}}$ is the nominal carrier frequency. The predicted geometric pseudorange-rate $\hat{\dot{\rho}}^i_{k}$ is computed from the relative GNSS-receiver velocity projected onto the LOS direction,
\begin{equation}
    \hat{\dot{\rho}}^i_{k}
    =
    \left(
        \hat{\mathbf{v}}^-_k - \mathbf{v}^{i}_{s,k}
    \right)^{\transpose}
    \mathbf{u}^i_{k},
    \label{eq:pred_prr_geom}
\end{equation}
where $\mathbf{v}^{i}_{s,k}$ is the broadcast GNSS satellite velocity; and $\mathbf{u}^i_{k} = \left(\mathbf{r}^{i}_{s,k} - \hat{\mathbf{r}}^{-}_{k}\right)\!/\norm{\mathbf{r}^{i}_{s,k} - \hat{\mathbf{r}}^{-}_{k}}$ is the unit vector along the receiver-to-GNSS LOS direction.

\subsubsection{Measurement vector and observation matrix}
If $n_k$ channels are active at epoch $k$, the measurement vector $\mathbf{z}_k$ is formed by stacking pseudorange errors and pseudorange-rate errors,
\begin{equation}
    \mathbf{z}_k
    =
    \begin{bmatrix}
        \delta \rho^1_{k},\cdots,\delta \rho^{n_k}_{k},
        \delta \dot{\rho}^1_{k},\cdots,\delta \dot{\rho}^{n_k}_{k}
    \end{bmatrix}^{\transpose}.
    \label{eq:meas_vector}
\end{equation}
The corresponding observation matrix for the $i$-th satellite is
\begin{equation}
    \mathbf{H}^i_{k}
    =
    \begin{bmatrix}
        -\left(\mathbf{u}^{i}_{k}\right)^{\transpose} & \mathbf{0}_{1\times3} 
        & \mathbf{0}_{1\times3} & 1 & 0 \\
        \mathbf{0}_{1\times3} & -\left(\mathbf{u}^{i}_{k}\right)^{\transpose} 
        & \mathbf{0}_{1\times3} & 0 & 1
    \end{bmatrix},
    \label{eq:Hi}
\end{equation}
and the full observation matrix is obtained by vertical stacking,
\begin{equation}
    \mathbf{H}_k
    =
    \begin{bmatrix}
        \left(\mathbf{H}_{k}^{1}\right)^{\transpose},
        \left(\mathbf{H}_{k}^{2}\right)^{\transpose},
        \cdots,
        \left(\mathbf{H}_{k}^{n_k}\right)^{\transpose}
    \end{bmatrix}^{\transpose}.
    \label{eq:Hstack}
\end{equation}

\subsubsection{EKF update}
Following the \textit{a priori} propagation shown by \eqref{eq:absolute_prediction} and \eqref{eq:cov_prediction}, the EKF \textit{a posteriori} measurement update is given by

\begin{align}
    \mathbf{y}_k &= \mathbf{z}_k - \mathbf{H}_k \delta \hat{\mathbf{X}}^-_k,
    \label{eq:innov} \\
    \mathbf{S}_k &= \mathbf{H}_k \mathbf{P}^-_k \mathbf{H}_k^{\transpose} + \mathbf{R}_k,
    \label{eq:innov_cov} \\
    \mathbf{K}_k &= \mathbf{P}^-_k \mathbf{H}_k^{\transpose}\mathbf{S}_k^{-1},
    \label{eq:kf_gain} \\
    \delta \hat{\mathbf{X}}^+_k &= \delta \hat{\mathbf{X}}^-_k + \mathbf{K}_k \mathbf{y}_k,
    \label{eq:err_update} \\
    \mathbf{P}^+_k &= \left(\mathbf{I}-\mathbf{K}_k\mathbf{H}_k\right)\mathbf{P}^-_k
\left(\mathbf{I}-\mathbf{K}_k\mathbf{H}_k\right)^{\transpose} 
+ \mathbf{K}_k\mathbf{R}_k\mathbf{K}_k^{\transpose},
    \label{eq:cov_update}
\end{align}
where $\mathbf{y}_k$ is the innovation; and $\mathbf{R}_k$ is the measurement noise covariance. The absolute navigation state is then corrected as
\begin{equation}
    \hat{\mathbf{X}}^+_{\text{GNSS},k} = \hat{\mathbf{X}}^-_{\text{GNSS},k} + \delta \hat{\mathbf{X}}^+_k,
    \label{eq:absolute_update}
\end{equation}

In the current implementation, the measurement noise covariance is initialised as diagonal with separate pseudorange and pseudorange-rate blocks,
\begin{equation}
    \mathbf{R}_0 = \mathrm{diag}\!\left(
        \left(\sigma^1_{\rho}\right)^2,\dots,\left(\sigma^{n_k}_{\rho}\right)^2,
        \left(\sigma^1_{\dot{\rho}}\right)^2,\dots,
        \left(\sigma^{n_k}_{\dot{\rho}}\right)^2
    \right).
    \label{eq:R_init}
\end{equation}
To account for time-varying signal conditions, an innovation-based adaptive measurement noise technique is implemented \citep{mohamed1999adaptive}. Every $N_R$ filter epochs, the covariance sub-blocks are refreshed using the $N_R$ most recent innovations as
\begin{align}
    \mathbf{R}^{\rho}_{m} &= \mathrm{diag}\left(
        \mathrm{clip}\left(
            \frac{1}{N_R c_{\rho}}\sum_{j=1}^{N_R}
            \mathbf{y}^{\rho}_{k-j+1}\odot\mathbf{y}^{\rho}_{k-j+1},\;
            r^{\rho}_{\min},\; r^{\rho}_{\max}
        \right)\right), \label{eq:R_rho} \\
    \mathbf{R}^{\dot{\rho}}_{m} &= \mathrm{diag}\left(
        \mathrm{clip}\left(
            \frac{1}{N_R c_{\dot{\rho}}}\sum_{j=1}^{N_R}
            \mathbf{y}^{\dot{\rho}}_{k-j+1}\odot\mathbf{y}^{\dot{\rho}}_{k-j+1},\;
            r^{\dot{\rho}}_{\min},\;r^{\dot{\rho}}_{\max}
        \right)\right), \label{eq:R_rhodot}
\end{align}
where $m = \lfloor k / N_R \rfloor$ is the epoch of measurement noise update; $c_{\rho}$ and $c_{\dot{\rho}}$ are sensitivity coefficients for the pseudorange and pseudorange-rate observables, respectively; and $r^{\rho}_{\min/\max}$ and $r^{\dot{\rho}}_{\min/\max}$ are the admissible lower and upper bounds. The full adapted covariance is assembled as

\begin{equation}
    \mathbf{R}_m = \mathrm{diag}\!\left(
        \mathbf{R}^{\rho}_{m},\;\mathbf{R}^{\dot{\rho}}_{m}
    \right).
    \label{eq:R_adapted}
\end{equation}

Algorithm~\ref{alg:ekf_update} summarises one receiver-side EKF epoch.

\begin{algorithm}[htbp]
\caption{GNSS EKF iteration at epoch $k$}
\label{alg:ekf_update}
\begin{algorithmic}[1]
\State Propagate $\delta\hat{\mathbf{X}}^-_k$ and $\mathbf{P}^-_k$ using \eqref{eq:absolute_prediction} and \eqref{eq:cov_prediction}
\For{each visible satellite $i$}
    \State Compute $\mathbf{r}^{i}_{s,k}$, $\mathbf{v}^{i}_{s,k}$, $c\delta t^i_{s,k}$, and $c\delta \dot{t}^{i}_{s,k}$
    \State Form $\delta \rho_{i,k}$
    \State Form $\breve{\dot{\rho}}^i_{k}$, $\hat{\dot{\rho}}^i_{k}$, and $\delta \dot{\rho}^i_{k}$
    \State Construct the LOS-based observation rows in \eqref{eq:Hi}
\EndFor
\State Stack all measurements into $\mathbf{z}_k$ and all observation rows into $\mathbf{H}_k$
\State Compute innovation $\mathbf{y}_k$, innovation covariance $\mathbf{S}_k$, and gain $\mathbf{K}_k$
\State Append $\mathbf{y}_k$ to the rolling innovation record
\If{$k$ is an $\mathbf{R}$ update epoch ($k \bmod N_R = 0$)}
    \State Refresh $\mathbf{R}^{\mathrm{upd}}_{\rho,m}$ and $\mathbf{R}^{\mathrm{upd}}_{\dot{\rho},m}$ from the last $N_R$ innovations using \eqref{eq:R_rho} and \eqref{eq:R_rhodot}
    \State Assemble $\mathbf{R}^{\mathrm{upd}}_m$ via \eqref{eq:R_adapted} and set $\mathbf{R}_k \gets \mathbf{R}^{\mathrm{upd}}_m$
\EndIf
\State Update the error state, covariance, and absolute state
\State Feed the updated navigation solution back to the code and carrier NCOs
\end{algorithmic}
\end{algorithm}

\subsection{Principle of orbital dynamic model}
\label{subsec:odm}

In the proposed algorithm, the ODM is used to provide a physically plausible short-term trajectory to stabilise the GNSS solution when the measurement update is temporarily weak or contaminated. The translational motion of a spacecraft is governed by the second-order differential equation \citep{dash2015bernese}
\begin{equation}
    \ddot{\mathbf{r}}(t)
    = \mathbf{a}_{\text{CG}}(t)+ \mathbf{a}_p(t) = 
    -\mu \frac{\mathbf{r}(t)}{\norm{\mathbf{r}(t)}^3} + \mathbf{a}_p(t),
    \label{eq:eom_general}
\end{equation}
where $\mathbf{r} \in \mathbb{R}^3$ is the position vector of the spacecraft, $\mathbf{a}_{\mathrm{CG}}(t)$ is the Earth's central gravity, $\mu = 3.98600 \times 10^{11}~\mathrm{km}^3\mathrm{s}^{-2}$ is the standard gravitational parameter of the Earth, and $\mathbf{a}_{p}(t)$ represents the collective effect of all perturbations. For notational brevity, the explicit time argument $(t)$ is dropped from all acceleration terms hereafter, i.e.\ $\mathbf{a}_{\mathrm{CG}} \equiv \mathbf{a}_{\mathrm{CG}}(t)$ and $\mathbf{a}_p \equiv \mathbf{a}_p(t)$, and this convention extends to all perturbation components introduced below. The perturbation term $\mathbf{a}_p$ can be expanded into
\begin{equation}
\mathbf{a}_p = \mathbf{a}_J + \mathbf{a}_{\mathrm{drag}} + \mathbf{a}_{3B} + \mathbf{a}_{\mathrm{SRP}} + \mathbf{a}_{\mathrm{misc}},
\label{eq:perturb}
\end{equation}
where $\mathbf{a}_{J}$ represents the $J_2$ and higher-order terms of the non-spherical geopotential; $\mathbf{a}_{\mathrm{drag}}$ is aerodynamic drag; $\mathbf{a}_{3\mathrm{B}}$ denotes the third-body attraction induced by other celestial bodies (e.g. the Sun and Moon); $\mathbf{a}_{\mathrm{SRP}}$ is solar radiation pressure (SRP); and $\mathbf{a}_{\mathrm{misc}}$ collects other small perturbations, including but not limited to general relativistic effects, tidal effects, and albedo effects \citep{montenbruck2002satellite}.

Building on \eqref{eq:perturb}, the ODM in this implementation is realised through the procedure summarised in Algorithm~\ref{alg:odm_propagation} \citep{mathlab2025odm}. Throughout the algorithm, left superscripts ${}^e(\cdot)$ and ${}^i(\cdot)$ denote quantities in the ECEF and Earth-Centred Inertial (ECI) frames, respectively, and the time-varying rotation operator $\mathcal{T}(t)$ maps a state between the ECEF and ECI frames at time $t$, with the subscript indicating the direction of transformation. The ODM is invoked every $N$ GNSS epochs. Given the initial navigation state, the ODM propagates the orbit of the LEO satellite forward over the interval $T_{\mathrm{ODM}}$ and returns the predicted position and velocity in the ECEF reference frame.

\begin{algorithm}[htbp]
\caption{Numerical ODM propagation at ODM update epoch $l$}
\label{alg:odm_propagation}
\begin{algorithmic}[1]
\State Set ${}^{e}\hat{\mathbf{r}}_{0}={}^e\hat{\mathbf{r}}^+_{l}$ and ${}^{e}\hat{\mathbf{v}}_{0}={}^e\hat{\mathbf{v}}^+_{l}$ in ECEF
\State Transform to the inertial frame: $({}^{i}\hat{\mathbf{r}}_{0},{}^{i}\hat{\mathbf{v}}_{0})=\mathcal{T}_{e\rightarrow i}(t_{l})({}^{e}\hat{\mathbf{r}}_{0},{}^{e}\hat{\mathbf{v}}_{0})$
\State Form Keplerian elements for propagator initialisation
\State Set the propagation horizon $[t_{l},\,t_{l+1}]$
\For{each internal numerical integration step $j$ within $T_{\mathrm{ODM}}$}
    \State $({}^{e}\hat{\mathbf{r}}_{j},{}^{e}\hat{\mathbf{v}}_{j})=\mathcal{T}_{i\rightarrow e}(t_{j})({}^{i}\hat{\mathbf{r}}_{j},{}^{i}\hat{\mathbf{v}}_{j})$
    \State Evaluate ECEF gravity ${}^{e}\mathbf{a}_{g,j} = {}^{e}\mathbf{a}_{\text{CG},j}+{}^{e}\mathbf{a}_{J,j}$ from the selected geopotential model
    \State Transform gravity to the inertial frame: ${}^{i}\mathbf{a}_{g,j}=\mathcal{T}_{e\rightarrow i}(t_{j})\;{}^{e}\mathbf{a}_{g,j}$
    \State Evaluate ${}^{i}\mathbf{a}_{\mathrm{drag},j}$, ${}^{i}\mathbf{a}_{3\mathrm{B},j}$, ${}^{i}\mathbf{a}_{\mathrm{SRP},j}$, and ${}^{i}\mathbf{a}_{\mathrm{misc},j}$
    \State Form ${}^{i}\mathbf{a}_{j} = {}^{i}\mathbf{a}_{g,j}+{}^{i}\mathbf{a}_{\mathrm{drag},j}+{}^{i}\mathbf{a}_{3\mathrm{B},j}+{}^{i}\mathbf{a}_{\mathrm{SRP},j}+{}^{i}\mathbf{a}_{\mathrm{misc},j}$
    \State Integrate ${}^{i}\dot{\mathbf{r}}_{j}={}^{i}\mathbf{v}_{j}$ and ${}^{i}\dot{\mathbf{v}}_{j}={}^{i}\mathbf{a}_{j}$ with a variable-step solver
\EndFor
\State $({}^e\hat{\mathbf{r}}^{-}_{\mathrm{ODM},l},{}^e\hat{\mathbf{v}}^{-}_{\mathrm{ODM},l})=\mathcal{T}_{i\rightarrow e}(t_{l+1})({}^i\hat{\mathbf{r}}_{l+1},{}^i\hat{\mathbf{v}}_{l+1})$
\State Output $\hat{\mathbf{X}}^{-}_{\mathrm{ODM},l+1}=\begin{bmatrix}{}^e\hat{\mathbf{r}}^{-}_{\mathrm{ODM},l+1},\; {}^e\hat{\mathbf{v}}^{-}_{\mathrm{ODM},l+1}\end{bmatrix}^{\transpose}$
\end{algorithmic}
\end{algorithm}

The processing framework adopts a low-fidelity approach to minimise computational cost, and the spacecraft is modelled as a standard 1U CubeSat, for which the detailed settings are listed in Table~\ref{tab:odm_parameters}.

\subsection{Adaptive orbit-aiding update}
\label{subsec:couplingfilter}
\subsubsection{ODM state residual}
In the proposed adaptive orbit-aiding update, the ODM state propagated in Section \autoref{subsec:odm} is treated as an auxiliary pseudo-measurement of the first six GNSS navigation states. The ODM residual relative to the GNSS EKF prediction is
\begin{equation}
    \mathbf{e}_{\text{ODM},l}
    =
    \hat{\mathbf{X}}_{\mathrm{ODM},l}
    -
    \mathbf{H}_{\mathrm{ODM}}\hat{\mathbf{X}}^+_{\text{GNSS},l},
    \label{eq:odm_residual}
\end{equation}
where
\begin{equation}
    \mathbf{H}_{\mathrm{ODM}}
    =
    \begin{bmatrix}
        \mathbf{I}_{6} & \mathbf{0}_{6\times5}
    \end{bmatrix}.
    \label{eq:Hodm}
\end{equation}

\subsubsection{GNSS stress scale}
The adaptive weighting rule uses the GNSS innovation history together with the instantaneous measurement geometry to detect stressed tracking conditions. Here, ``stressed'' refers to epochs at which the GNSS measurements become less consistent with the filter prediction than under nominal operation, for example because of weaker signals, poorer geometry, or tracking disturbances. 

Let the GNSS innovation vector at epoch $l$ be $\mathbf{y}_l$ from \eqref{eq:innov}. The RMS values of the pseudorange and pseudorange-rate innovations are used as compact scalar indicators of the aggregate innovation magnitude in the code and Doppler domains, respectively. The corresponding RMS quantities at epoch $l$ are
\begin{align}
    \delta \rho_{\text{RMS},l}
    &=
    \sqrt{
        \frac{1}{n_l}
        \sum_{i=1}^{n_l}
        \left(\delta \rho_{l}^{i}\right)^2
    },
    \label{eq:rms_pr} \\
    \delta \dot{\rho}_{\text{RMS},l}
    &=
    \sqrt{
        \frac{1}{n_l}
        \sum_{i=1}^{n_l}
        \left(\delta \dot{\rho}_{l}^{i}\right)^2
    }.
    \label{eq:rms_prr}
\end{align}
To account explicitly for geometry quality, let $\mathbf{G}_l\in\mathbb{R}^{n_l\times 4}$ denote the pseudorange-geometry submatrix formed from the pseudorange rows of $\mathbf{H}_l$ and the state columns corresponding to receiver position and clock bias. The corresponding GDOP is
\begin{equation}
    \mathrm{GDOP}_l
    =
    \sqrt{\mathrm{tr}\!\left(\left(\mathbf{G}_l^{\transpose}\mathbf{G}_l\right)^{-1}\right)},
    \label{eq:gdop}
\end{equation}
and the implemented geometry-penalty factor is
\begin{equation}
    g_l
    =
    \begin{cases}
        1, & \mathrm{GDOP}_l \le g_{\mathrm{ref}}, \\
        \min\!\left[\left(\dfrac{\mathrm{GDOP}_l}{g_{\mathrm{ref}}}\right)^{c_g},\, g_{\max}\right], & \mathrm{GDOP}_l > g_{\mathrm{ref}},
    \end{cases}
    \label{eq:gdop_penalty}
\end{equation}
where $g_{\mathrm{ref}}$ is the GDOP normalisation reference level; $c_g$ is the geometry sensitivity coefficient; and $g_{\max}$ is the admissible upper bound on the geometry penalty. These terms are subsequently combined into a GNSS stress scale
\begin{equation}
    s_{\mathrm{GNSS},l}
    =
    \mathrm{clip}\!\left(
        \left[
            1
            +
            c_s
            \left(
                \frac{\delta \rho_{\text{RMS},l}}{\rho_{\mathrm{ref}}}
                +
                \frac{\delta \dot{\rho}_{\text{RMS},l}}{\dot{\rho}_{\mathrm{ref}}}
            \right),\;
        \right]
        g_l,\;
        1,\;
        s_{\max}
    \right),
    \label{eq:gnss_stress}
\end{equation}
where $c_s$ is the RMS stress sensitivity coefficient; $\rho_{\mathrm{ref}}$ and $\dot{\rho}_{\mathrm{ref}}$ are the normalisation reference values for $\delta \rho_{\text{RMS},l}$ and $\delta \dot{\rho}_{\text{RMS},l}$, respectively; and $s_{\max}$ is the admissible upper bound. In the implementation, if fewer than four pseudorange rows are available to form $\mathbf{G}_l$, or if $\mathbf{G}_l^{\transpose}\mathbf{G}_l$ is ill-conditioned, the geometry penalty is set directly to $g_{\max}$. If valid innovation RMS values are unavailable, the stress scale is set to $s_{\max}$. These conventions ensure that the adaptive weighting rule assigns the ODM pseudo-measurement maximal authority during GNSS outages or severely degraded geometry.

\subsubsection{Adaptive ODM covariance assignment}
The ODM pseudo-measurement covariance is adjusted at each scheduled fusion epoch so that the dynamic-model confidence remains responsive to both the current ODM--GNSS agreement and the inferred GNSS stress level.

Position and velocity residuals are treated separately,
\begin{equation}
    \mathbf{e}_{\text{ODM},l}
    =
    \begin{bmatrix}
        \mathbf{e}^{r}_{l} \\
        \mathbf{e}^{v}_{l}
    \end{bmatrix},
    \qquad
    \mathbf{e}^{r}_{l},\mathbf{e}^{v}_{l}\in\mathbb{R}^{3}.
    \label{eq:residual_partition}
\end{equation}
The adaptive standard deviations (STDs) are then assigned component-wise as
\begin{align}
    \bm{\sigma}^{r}_{l}
    &=
    \mathrm{clip}\!\left(
        \bm{\sigma}^{r}_{0}
        \odot
        \left[
            \bm{1}
            +
            c_r
            \frac{\abs{\mathbf{e}^{r}_{l}}}{\mathbf{r}_{\mathrm{ref}}}
        \right]
        \oslash
        s_{\mathrm{GNSS},l},
        \bm{\sigma}^{r}_{\min},
        \bm{\sigma}^{r}_{\max}
    \right),
    \label{eq:sigma_r} \\
    \bm{\sigma}^{v}_{l}
    &=
    \mathrm{clip}\!\left(
        \bm{\sigma}^{v}_{0}
        \odot
        \left[
            \bm{1}
            +
            c_v
            \frac{\abs{\mathbf{e}^{v}_{l}}}{\mathbf{v}_{\mathrm{ref}}}
        \right]
        \oslash
        s_{\mathrm{GNSS},l},
        \bm{\sigma}^{v}_{\min},
        \bm{\sigma}^{v}_{\max}
    \right),
    \label{eq:sigma_v}
\end{align}
where $\bm{\sigma}^{r}_{0},\bm{\sigma}^{v}_{0}\in\mathbb{R}^3$ are the baseline position and velocity STD vectors, respectively; $c_r$ and $c_v$ are residual sensitivity coefficients for position and velocity, respectively; $\mathbf{r}_{\mathrm{ref}},\mathbf{v}_{\mathrm{ref}}\in\mathbb{R}^3$ are normalisation reference vectors for $\abs{\mathbf{e}^{r}_{l}}$ and $\abs{\mathbf{e}^{v}_{l}}$, respectively; and $\bm{\sigma}^{r}_{\min/\max}, \bm{\sigma}^{v}_{\min/\max}\in\mathbb{R}^3$ are the admissible lower and upper bounds applied component-wise. As $s_{\mathrm{GNSS},l}$ enters \eqref{eq:sigma_r} and \eqref{eq:sigma_v} as a divisor, a larger stress drives the assigned STDs downward. If valid innovation RMS values are unavailable, the implementation assigns $\bm{\sigma}^{r}_l=\bm{\sigma}^{r}_{\min}$ and $\bm{\sigma}^{v}_l=\bm{\sigma}^{v}_{\min}$ so that the ODM receives maximal authority during GNSS outages.

With this method, larger ODM residual components increase the assigned STDs and therefore reduce the authority of the corresponding ODM components, whereas a larger GNSS stress $s_{\mathrm{GNSS},l}$ decreases the assigned STDs and therefore raises the authority of the corresponding ODM components when the GNSS solution is less reliable. The full ODM covariance is therefore
\begin{equation}
    \mathbf{R}_{\mathrm{ODM},l}
    =
    \diag\!\left(
        \bm{\sigma}^{r}_{l}\odot\bm{\sigma}^{r}_{l},\;
        \bm{\sigma}^{v}_{l}\odot\bm{\sigma}^{v}_{l}
    \right).
    \label{eq:R_odm}
\end{equation}

\subsubsection{Residual gating}
Before applying the ODM correction, the residual is normalised by the adaptive STDs,
\begin{equation}
    \bm{\eta}_l
    =
    \abs{\mathbf{e}_{\text{ODM},l}}
    \oslash
    \bm{\sigma}_{l},
    \qquad
    \bm{\sigma}_l
    =
    \begin{bmatrix}
        \bm{\sigma}^{r}_{l} \\
        \bm{\sigma}^{v}_{l}
    \end{bmatrix}.
    \label{eq:eta}
\end{equation}
where $\bm{\eta}_l\in\mathbb{R}^6$ is the absolute residual in units of the adaptive STDs. Let the gate-threshold vector be
\begin{equation}
    \bm{\gamma}
    =
    \begin{bmatrix}
        \gamma_r \bm{1}_3 \\
        \gamma_v \bm{1}_3
    \end{bmatrix},
    \label{eq:gate_threshold}
\end{equation}
where $\gamma_r$ and $\gamma_v$ are the position and velocity gate thresholds, respectively. The ODM update is rejected if
\begin{equation}
    \exists i \in \{1,\dots,6\}
    \ \text{such that} \
    \eta_{i,l} > \gamma_i,
    \label{eq:gate_condition}
\end{equation}
in which case

\begin{equation}
    \hat{\mathbf{X}}^{+}_{l}
    =
    \hat{\mathbf{X}}^{+}_{\mathrm{GNSS},l}.
    \label{eq:gated_skip}
\end{equation}
This prevents the ODM from pulling the navigation solution towards an evidently inconsistent state if the ODM itself enters a fault state.

\subsubsection{Adaptive pseudo-measurement update}
If the residual passes the gate test, the ODM is fused through a KF-style correction,
\begin{align}
    \mathbf{K}_{\mathrm{ODM},l}
    &=
    \mathbf{P}^{+}_{\mathrm{GNSS},l}
    \mathbf{H}_{\mathrm{ODM}}^{\transpose}
    \left(
        \mathbf{H}_{\mathrm{ODM}}
        \mathbf{P}^{+}_{\mathrm{GNSS},l}
        \mathbf{H}_{\mathrm{ODM}}^{\transpose}
        +
        \mathbf{R}_{\mathrm{ODM},l}
    \right)^{-1},
    \label{eq:adaptive_gain} \\
    \delta \hat{\mathbf{X}}_{\mathrm{ODM},l}
    &=
    \mathbf{K}_{\mathrm{ODM},l} \mathbf{e}_{\mathrm{ODM},l},
    \label{eq:adaptive_correction} \\
    \hat{\mathbf{X}}^{+}_{l}
    &=
    \hat{\mathbf{X}}^{+}_{\mathrm{GNSS},l}
    +
    \delta \hat{\mathbf{X}}_{\mathrm{ODM},l}.
    \label{eq:adaptive_state_update}
\end{align}
The error-state covariance is updated in Joseph form,
\begin{equation}
    \mathbf{P}^{+}_{l}
    =
    \left(
        \mathbf{I}
        -
        \mathbf{K}_{\mathrm{ODM},l}\mathbf{H}_{\mathrm{ODM}}
    \right)
    \mathbf{P}^{+}_{\mathrm{GNSS},l}
    \left(
        \mathbf{I}
        -
        \mathbf{K}_{\mathrm{ODM},l}\mathbf{H}_{\mathrm{ODM}}
    \right)^{\transpose}
    +
    \mathbf{K}_{\mathrm{ODM},l}
    \mathbf{R}_{\mathrm{ODM},l}
    \left(
        \mathbf{K}_{\mathrm{ODM},l}
    \right)^{\transpose},
    \label{eq:joseph}
\end{equation}
which preserves the symmetry and positive semi-definiteness of the covariance.

Algorithm~\ref{alg:adaptive_odm} gives the full procedure of the adaptive orbit-aiding update.

\begin{algorithm}[htbp]
\caption{Adaptive orbit-aiding update at ODM update epoch $l$}
\label{alg:adaptive_odm}
\begin{algorithmic}[1]
\State Read the propagated ODM state $\hat{\mathbf{X}}_{\mathrm{ODM},l}$
\State Form the residual $\mathbf{e}_{\mathrm{ODM},l}$ using \eqref{eq:odm_residual}
\State Compute $\delta \rho_{\text{RMS},l}$ and $\delta \dot{\rho}_{\text{RMS},l}$ from GNSS innovations
\State Compute the GDOP penalty $g_l$ from the GNSS observation geometry
\State Compute the GNSS stress scale $s_{\mathrm{GNSS},l}$
\State Assign $\bm{\sigma}^{r}_{l}$ and $\bm{\sigma}^{v}_{l}$ using \eqref{eq:sigma_r} and \eqref{eq:sigma_v}
\State Form the normalised residual vector $\bm{\eta}_l$
\If{any component of $\bm{\eta}_l$ exceeds the gate threshold $\bm{\gamma}$}
    \State Reject the ODM update and keep the GNSS solution
\Else
    \State Form $\mathbf{R}_{\mathrm{ODM},l}$
    \State Compute $\mathbf{K}_{\mathrm{ODM},l}$
    \State Update the state $\hat{\mathbf{X}}^{+}_{l}$ and covariance $\mathbf{P}^{+}_{l}$ using \eqref{eq:adaptive_state_update} and \eqref{eq:joseph}
\EndIf
\end{algorithmic}
\end{algorithm}

\section{Results and analysis}
\label{sec:results}

This section presents the experimental design and performance assessment of adaptive OAVT. The evaluation is organised to examine baseline performance under nominal tracking conditions and robustness under degraded measurements and signal outages. The receiver configuration and parameter settings are first summarised in \autoref{subsec:results_setup}; the nominal-condition results are presented in \autoref{subsec:results_nominal}; and the degraded-scenario results are given in \autoref{subsec:results_degraded}.

\subsection{Receiver configuration and parameter tuning}
\label{subsec:results_setup}
The main receiver settings used to generate the reported results are summarised in Table~\ref{tab:results_setup_parameters}. These parameters define the common processing configuration adopted for the OAVT experiments, including the tracking update interval, navigation-state dimension, adaptive measurement-noise update strategy, and the baseline adaptive ODM tuning. The detailed ODM propagation settings are listed separately in Table~\ref{tab:odm_parameters}.

\begin{table}[htbp]
\caption{Algorithm configuration and parameters}
\label{tab:results_setup_parameters}
\centering
\begin{tabular}{|p{0.34\linewidth}|p{0.56\linewidth}|}
\hline
\textbf{Parameter} & \textbf{Setting} \\
\hline
EKF update interval & $T = 1$ ms \\
\hline
Process noise& $\mathbf{Q}=\diag\!\left(5\times10^{-2}\bm{1}_3,\ 5\times10^{-3}\bm{1}_3,\ 1\times10^{-2}\bm{1}_3,\ 5\times10^{-3},\ 1\times10^{-2}\right)$ \\
\hline
Initial measurement noise & $\mathbf{R}_0=\diag\!\left(2\times10^{-1}\bm{1}_{n_k},\ 2\times10^{-2}\bm{1}_{n_k}\right)$\\
\hline
Adaptive $\mathbf{R}$ update interval & $N_R = 1$ s \\
\hline
Adaptive $\mathbf{R}$ sensitivity coefficients & $c_{\rho} = 10$, $c_{\dot{\rho}} = 10$ \\
\hline
Adaptive $\mathbf{R}$ bounds & $r^{\rho}_{\min/\max} = [0.01, 1000]$; $r^{\dot{\rho}}_{\min/\max} = [0.01, 400]$ \\
\hline
ODM update interval & $T_{\mathrm{ODM}} = 0.1$ s \\
\hline
GDOP penalty tuning & $c_g = 0.5$, $g_{\mathrm{ref}} = 2.5$, $g_{\max} = 4$ \\
\hline
GNSS stress tuning & $c_s = 0.35$, $\rho_{\mathrm{ref}} = 20$ m, $\dot{\rho}_{\mathrm{ref}} = 2$ m/s, $s_{\max}=4$ \\
\hline
$\bm{\sigma}^{r/v}$ sensitivity coefficients & $c_r = 0.75$, $c_v = 0.75$ \\
\hline
$\bm{\sigma}^{r/v}$ reference scales & $\mathbf{r}_{\mathrm{ref}} = [20,\ 20,\ 30]^{\transpose}$ m, $\mathbf{v}_{\mathrm{ref}} = [0.50,\ 0.50,\ 0.75]^{\transpose}$ m/s \\
\hline
$\bm{\sigma}^{r/v}$ baseline setting & $\bm{\sigma}^{r}_{0} = [0.15,\ 0.15,\ 0.25]^{\transpose}$ m, $\bm{\sigma}^{v}_{0} = [0.005,\ 0.005,\ 0.0075]^{\transpose}$ m/s \\
\hline
$\bm{\sigma}^{r/v}$ bounds & $\bm{\sigma}^{r}_{\min/\max} = [0.02,0.02,0.04]$–$[250,250,350]^{\mathrm{T}}$ m;\newline $\bm{\sigma}^{v}_{\min/\max} = [0.0005,0.0005,0.0008]$–$[10,10,12]^{\mathrm{T}}$ m/s \\
\hline
Residual gate thresholds & $\gamma_r = 100$, $\gamma_v = 100$ \\
\hline
\end{tabular}
\end{table}

\subsection{Performance under nominal conditions}
\label{subsec:results_nominal}

The nominal-condition evaluation is conducted using Global Positioning System (GPS) L1 signals generated by a Spirent GSS7000 GNSS simulator and converted into IF data by a LABSAT3 Wideband front-end. The dataset represents a LEO spaceborne receiver at an altitude of 515 km, which is typical of CubeSat and LEO-PNT mission profiles. The onboard oscillator is modelled as a typical TCXO, and the simulated GNSS measurements retain consistently high signal-to-noise ratios without measurement degradation or signal outages. For the excerpt reported here, the data are processed over 200 s, during which 10 GNSS satellites are acquired. In this subsection, the STL solution produced by the GSRx Software GNSS Receiver \citep{soderholm2022fgigsrx} is used as the benchmark reference. The STL, VTL, and OAVT results are generated from the same simulated data, and their loop and navigation settings are tuned to be identical whenever the receiver architectures permit. Accordingly, the reported differences can be interpreted primarily as the effect of the receiver architecture rather than as a consequence of inconsistent parameterisation.

\begin{figure}[htbp]
    \centering
    \begin{subfigure}[b]{0.45\linewidth}
        \centering
        \includegraphics[width=\linewidth]{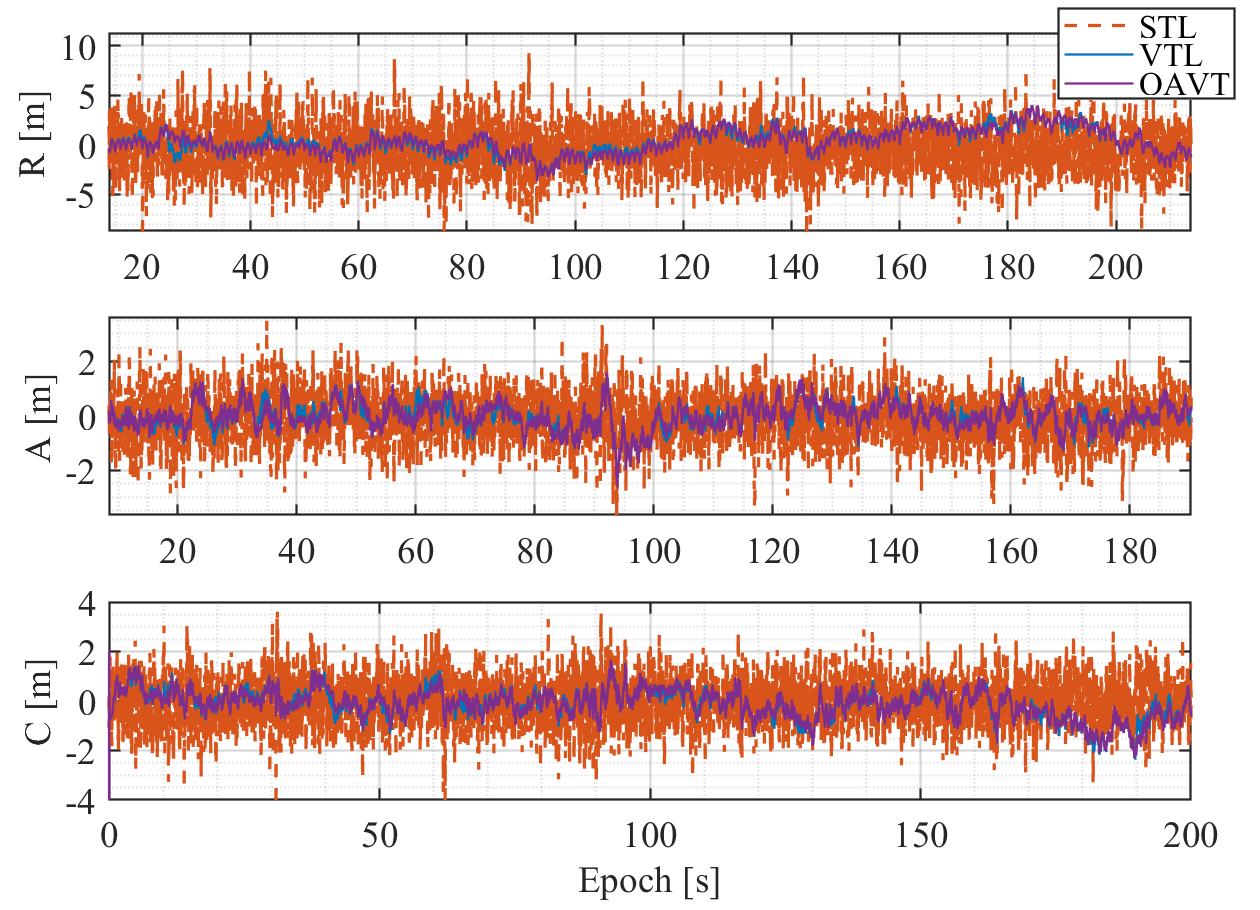}
        \subcaption{}
        \label{fig:favPosRAC}
    \end{subfigure}
    \hfill
    \begin{subfigure}[b]{0.45\linewidth}
        \centering
        \includegraphics[width=\linewidth]{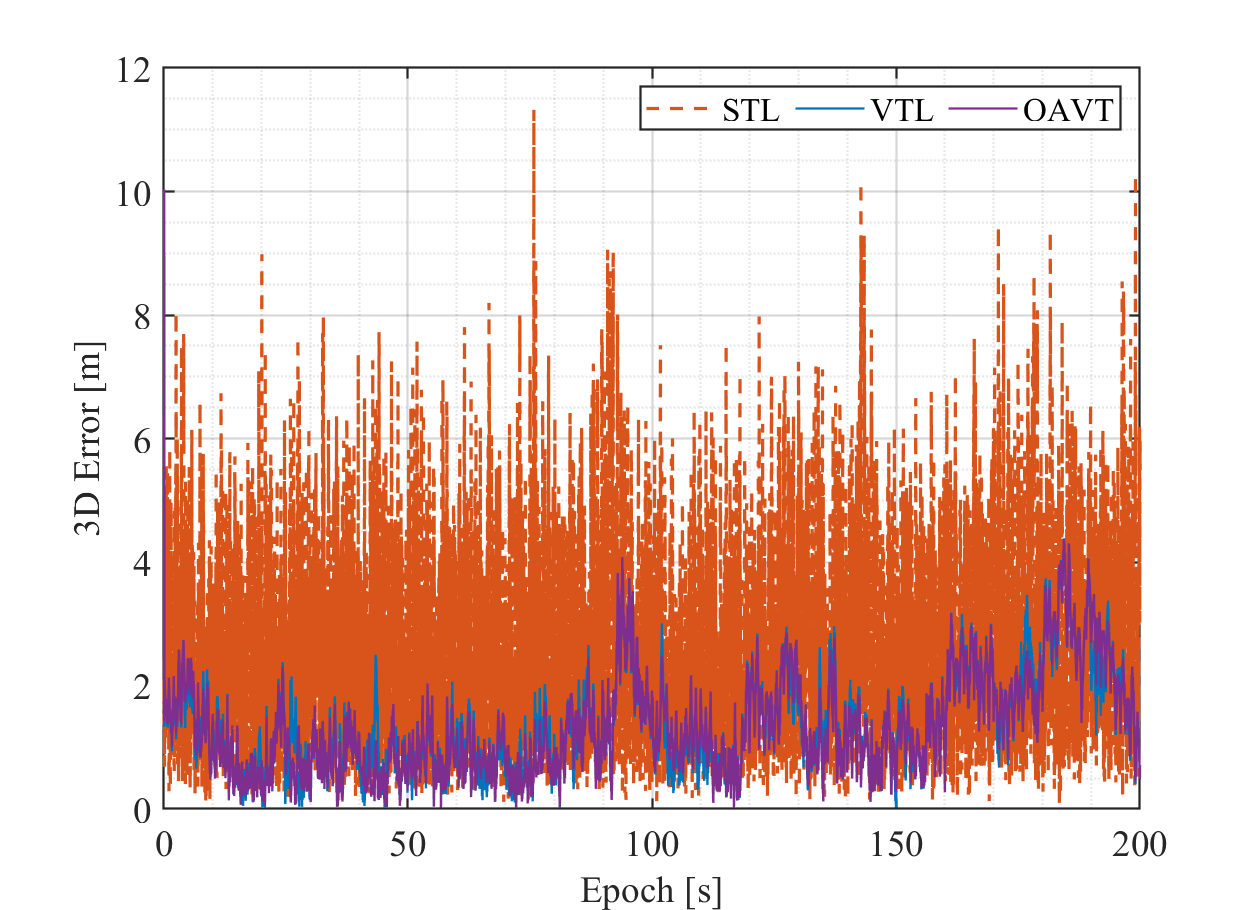}
        \subcaption{}
        \label{fig:favPos3D}
    \end{subfigure}
    \caption{Position errors of STL, VTL, and OAVT under nominal conditions (\textbf{a}) in radial, along-track, and cross-track components; (\textbf{b}) in 3D.}
    \label{fig:favPosCombined}
\end{figure}

\begin{figure}[htbp]
    \centering
    \begin{subfigure}[b]{0.45\linewidth}
        \centering
        \includegraphics[width=\linewidth]{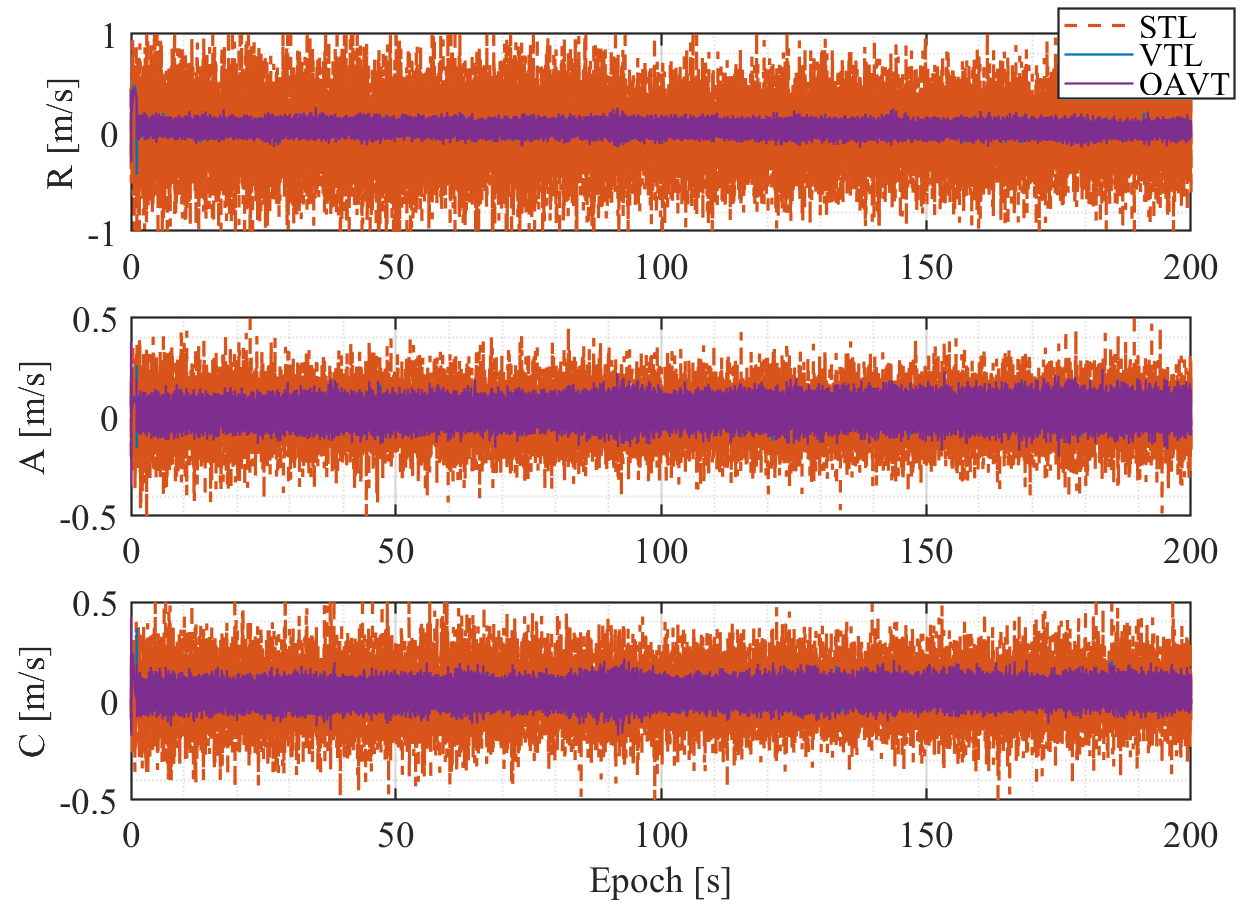}
        \subcaption{}
        \label{fig:favVelRAC}
    \end{subfigure}
    \hfill
    \begin{subfigure}[b]{0.45\linewidth}
        \centering
        \includegraphics[width=\linewidth]{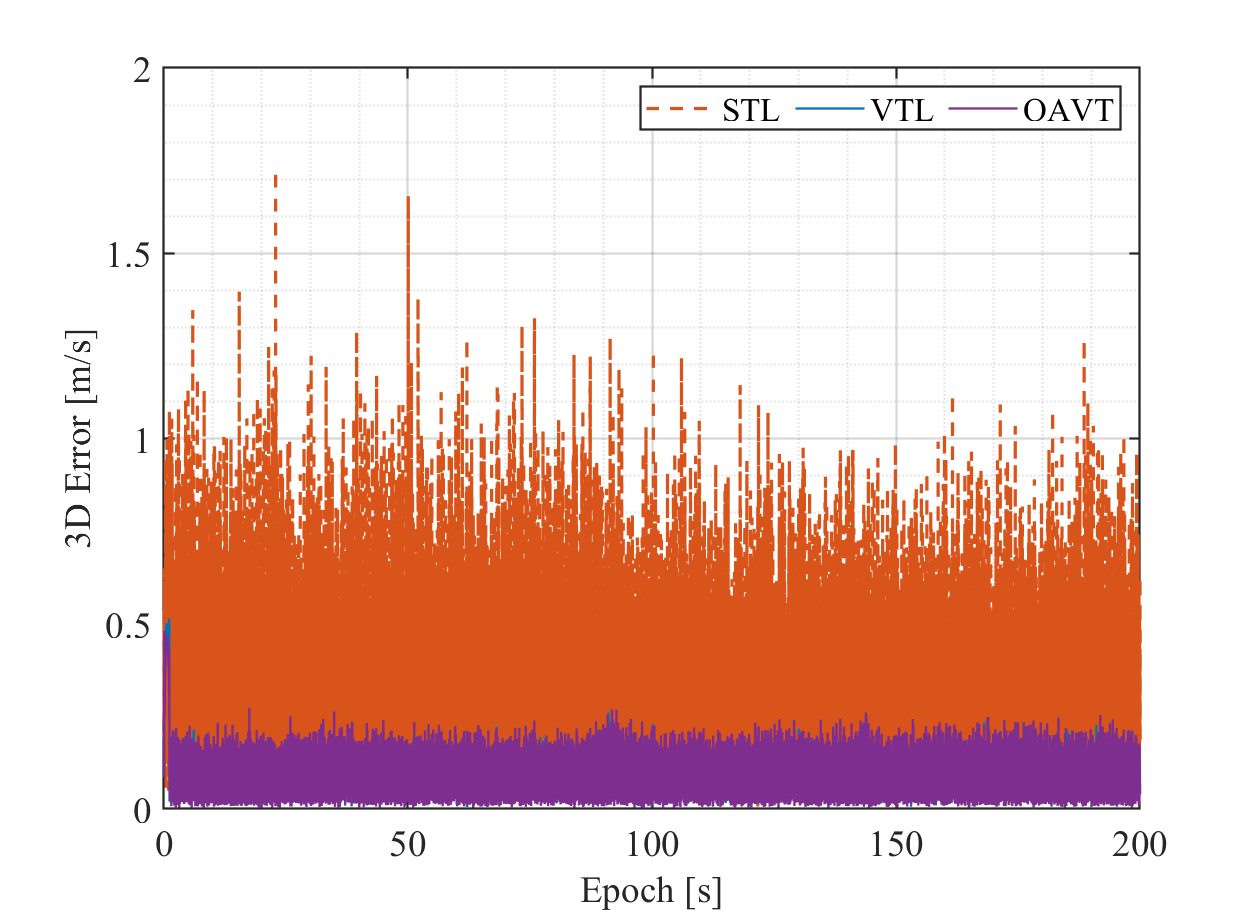}
        \subcaption{}
        \label{fig:favVel3D}
    \end{subfigure}
    \caption{Velocity errors of STL, VTL, and OAVT under nominal conditions (\textbf{a}) in radial, along-track, and cross-track components; (\textbf{b}) in 3D.}
    \label{fig:favVelCombined}
\end{figure}

The position error results are presented in \autoref{fig:favPosCombined}. Across the radial (R), along-track (A), and cross-track (C) components, both the VTL and OAVT solutions exhibit much smaller scatter than STL. This behaviour is consistent with the root-mean-square errors (RMSEs) presented in \autoref{tab:fav_rmse}, indicating that the vector tracking architecture provides substantially higher accuracy. The three-dimensional (3D) position RMSE is reduced from 2.631 m for STL to 1.082 m for VTL and 1.081 m for OAVT, corresponding to an improvement of approximately 58.9\% for both vector-based methods. At the component level, the largest reduction relative to STL is observed in the radial position error, where the RMSE decreases from 2.325 m to 0.837 m for VTL and 0.852 m for OAVT. The persistently poorer accuracy in the radial direction relative to the other two directions is consistent with other studies \citep{chen2017sf,yang2016ckf}. This can be explained by the fact that, analogous to altitude estimation in terrestrial scenarios, the GNSS vertical dilution of precision (VDOP) is naturally higher than the horizontal dilution of precision (HDOP), whereas the dynamic model embedded in the vector tracking filter helps constrain the error. In the along-track and cross-track directions, OAVT is marginally better than VTL, although the difference between the two remains small. This near-overlap between VTL and OAVT indicates that, under nominal conditions with consistent GNSS measurement quality, the adaptive orbit aid remains active but non-disruptive to the solution delivered by the core VTL estimator.

\begin{figure}[htbp]
    \centering
    \begin{subfigure}[b]{0.45\linewidth}
        \centering
        \includegraphics[width=\linewidth]{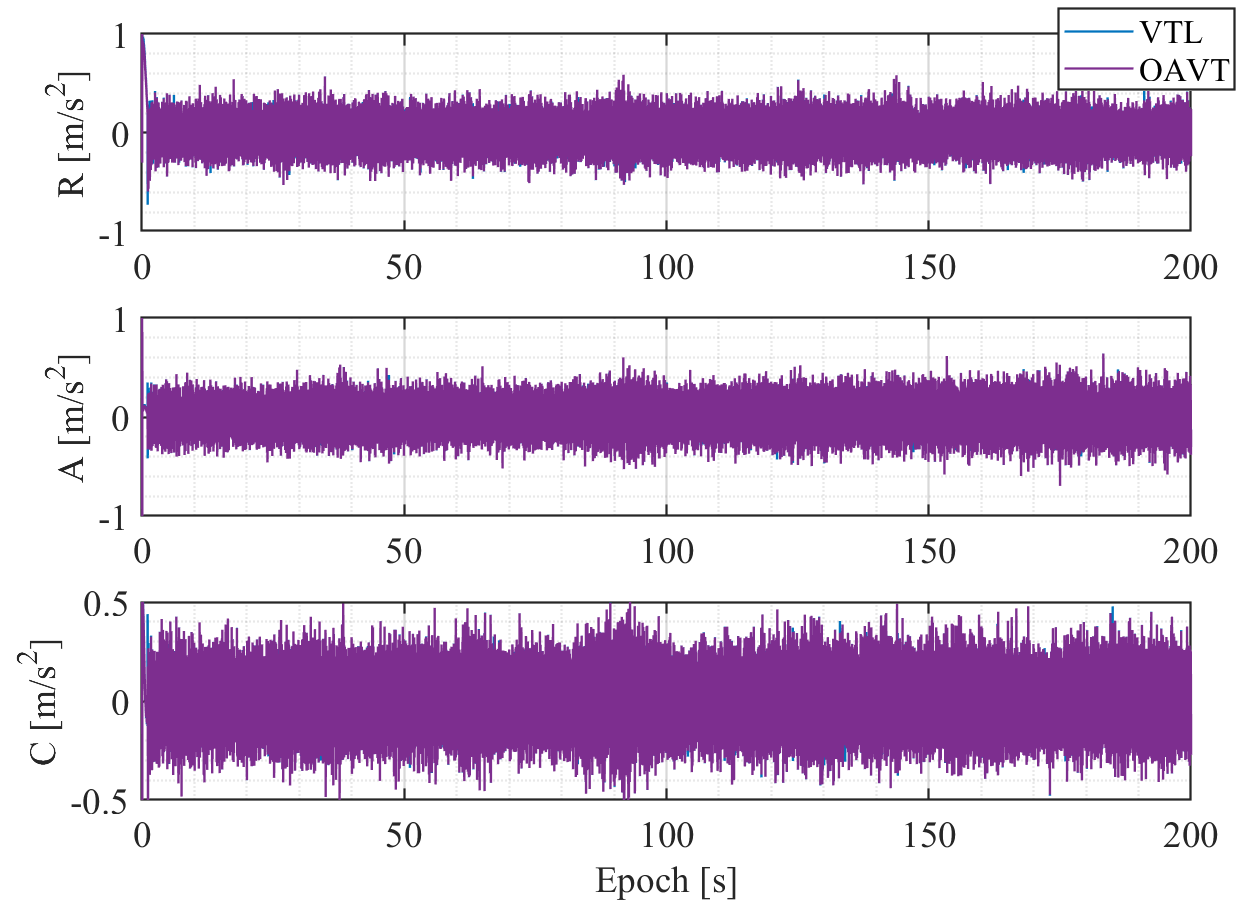}
        \subcaption{}
        \label{fig:favAccRAC}
    \end{subfigure}
    \hfill
    \begin{subfigure}[b]{0.45\linewidth}
        \centering
        \includegraphics[width=\linewidth]{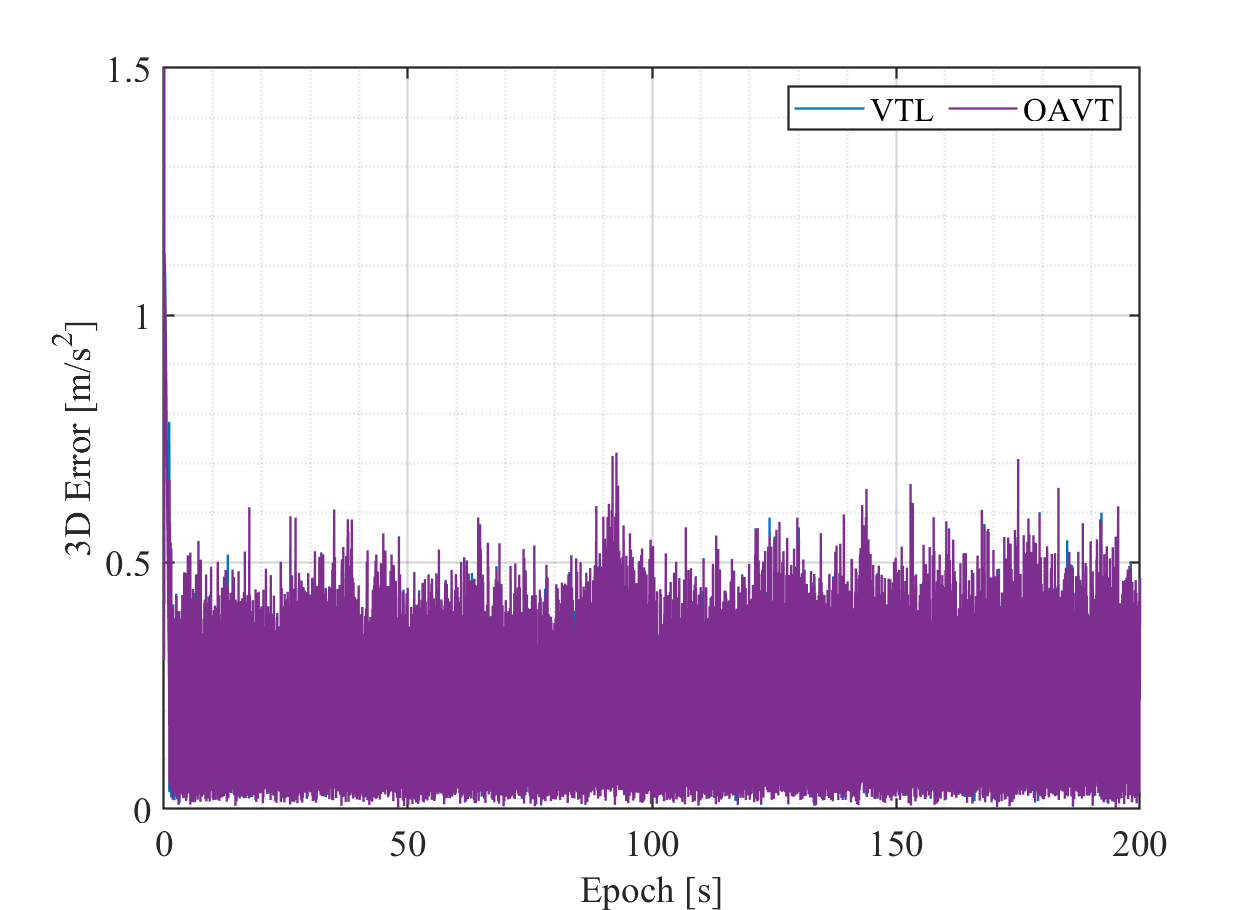}
        \subcaption{}
        \label{fig:favAcc3D}
    \end{subfigure}
    \caption{Acceleration errors of VTL and OAVT under nominal conditions (\textbf{a}) in radial, along-track, and cross-track components; (\textbf{b}) in 3D.}
    \label{fig:favAccCombined}
\end{figure}

A similar pattern is observed in velocity estimation in \autoref{fig:favVelCombined}. The STL velocity errors retain visibly broader fluctuations in all three RAC components, whereas the VTL and OAVT solutions exhibit significantly lower variance. The 3D velocity RMSE decreases from 0.3985 m/s for STL to 0.1029 m/s for VTL and 0.1023 m/s for OAVT, corresponding to improvements of 74.2\% and 74.3\%, respectively. The radial component again shows the most pronounced gain, with the RMSE falling from 0.3479 m/s for STL to 0.0675 m/s for VTL and 0.0667 m/s for OAVT. These results confirm that the principal gain in favourable conditions comes from the vector tracking architecture itself, owing to its exploitation of the coupling between observable estimation and dynamics prediction.

Since the STL configuration used does not provide acceleration estimates, only the VTL and OAVT acceleration errors are compared in \autoref{fig:favAccCombined}. Apart from the initial transient and a few local deviations, the two solutions again almost fully overlap, with 3D RMSE values of 0.2244~m/s$^2$ for VTL and 0.2233~m/s$^2$ for OAVT. The initial transient spikes observed in the velocity and acceleration errors are attributed to the large initial state covariance combined with unconverged receiver clock bias estimates, which together yield an elevated Kalman gain and a transient overshoot before the filter reaches steady-state performance. With the assistance of the ODM constraints, the OAVT solution converges slightly faster than the VTL solution.

\begin{table}[htbp]
\caption{RMSE comparison among STL, VTL, and OAVT under nominal conditions.}
\label{tab:fav_rmse}
\centering
\small
\setlength{\tabcolsep}{4pt}
\renewcommand{\arraystretch}{1.1}
\begin{tabular}{|>{\centering\arraybackslash}p{0.18\linewidth}|>{\centering\arraybackslash}p{0.15\linewidth}|>{\centering\arraybackslash}p{0.12\linewidth}|>{\centering\arraybackslash}p{0.12\linewidth}|>{\centering\arraybackslash}p{0.12\linewidth}|}
\hline
\textbf{Quantity} & \textbf{Component} & \textbf{STL} & \textbf{VTL} & \textbf{OAVT} \\
\hline
\multirow{4}{*}{Position error [m]}
 & Radial      & 2.325 & 0.837 (\textcolor{green!60!black}{64.0\%}) & 0.852 (\textcolor{green!60!black}{63.4\%})\\
\cline{2-5}
 & Along-track & 0.850 & 0.422 (\textcolor{green!60!black}{50.4\%}) & 0.402 (\textcolor{green!60!black}{52.7\%}) \\
\cline{2-5}
 & Cross-track & 0.891 & 0.540 (\textcolor{green!60!black}{39.4\%}) & 0.532 (\textcolor{green!60!black}{40.3\%}) \\
\cline{2-5}
 & 3D          & 2.631 & 1.082 (\textcolor{green!60!black}{58.9\%}) & 1.081 (\textcolor{green!60!black}{58.9\%}) \\
\hline
\multirow{4}{*}{Velocity error [m/s]}
 & Radial      & 0.3479 & 0.0675 (\textcolor{green!60!black}{80.6\%}) & 0.0667 (\textcolor{green!60!black}{80.8\%}) \\
\cline{2-5}
 & Along-track & 0.1285 & 0.0528 (\textcolor{green!60!black}{58.9\%}) & 0.0527 (\textcolor{green!60!black}{58.9\%}) \\
\cline{2-5}
 & Cross-track & 0.1459 & 0.0570 (\textcolor{green!60!black}{60.9\%}) & 0.0569 (\textcolor{green!60!black}{61.0\%}) \\
\cline{2-5}
 & 3D          & 0.3985 & 0.1029 (\textcolor{green!60!black}{74.2\%}) & 0.1023 (\textcolor{green!60!black}{74.3\%}) \\
\hline
\multirow{4}{*}{Acceleration error [m/s$^2$]}
 & Radial      & -- & 0.1342 & 0.1330 \\
\cline{2-5}
 & Along-track & -- & 0.1361 & 0.1359 \\
\cline{2-5}
 & Cross-track & -- & 0.1175 & 0.1172 \\
\cline{2-5}
 & 3D          & -- & 0.2244 & 0.2233 \\
\hline
\end{tabular}
\end{table}

\begin{figure}[htbp]
    \centering
    \includegraphics[width=0.6\linewidth]{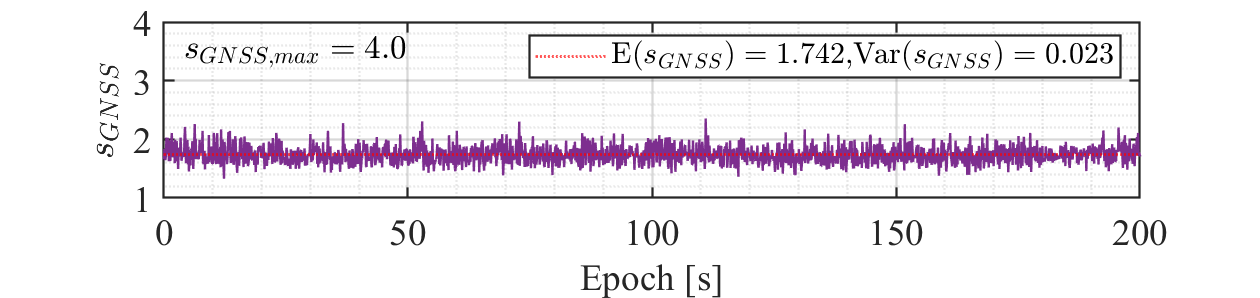}
    \caption{GNSS stress of OAVT under nominal conditions.}
    \label{fig:favGnssScale}
\end{figure}

\begin{figure}[htbp]
    \centering
    \includegraphics[width=0.5\linewidth]{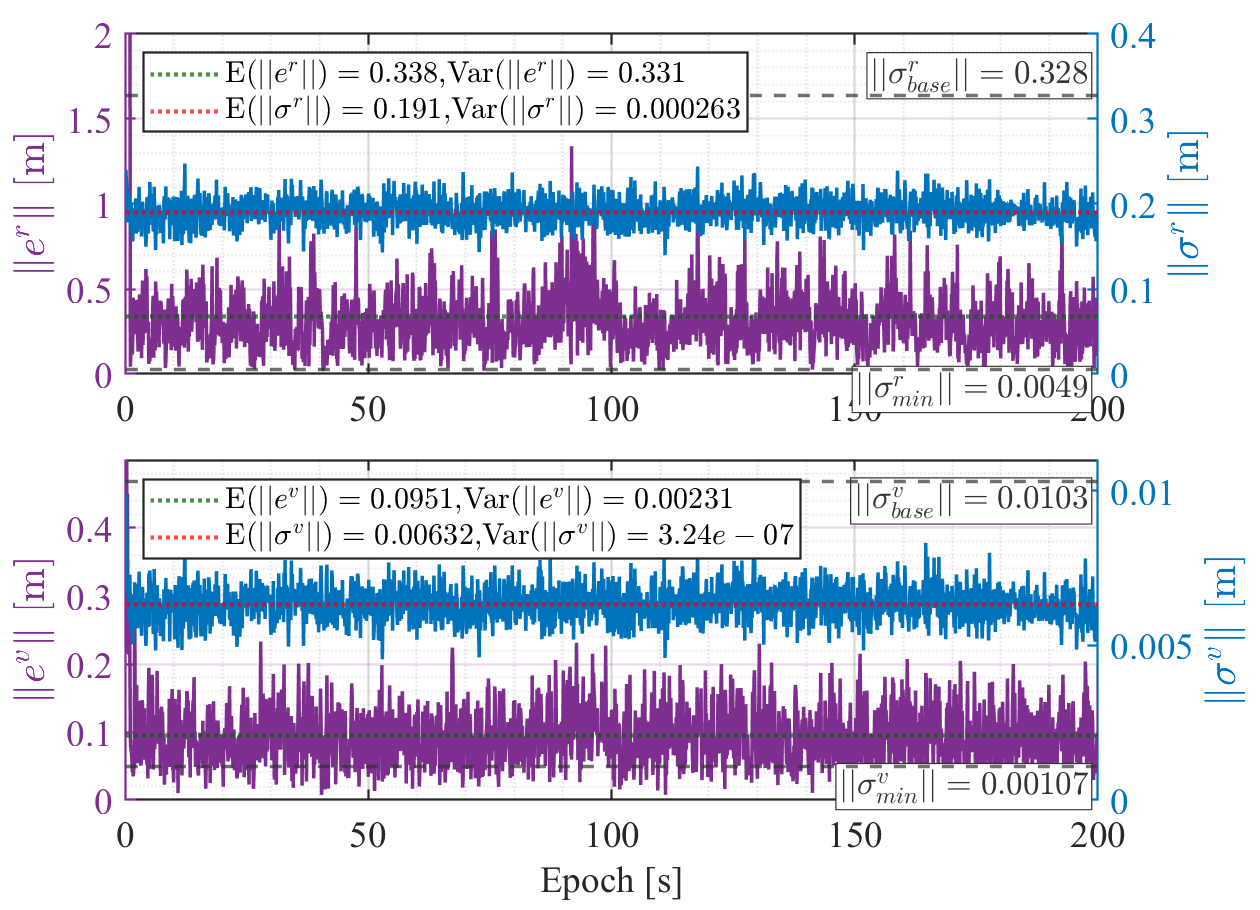}
    \caption{The state residual and adaptive covariance magnitudes of the ODM pseudo-measurements under nominal conditions.}
    \label{fig:favRsigma}
\end{figure}

\begin{figure}[htbp]
    \centering
    \includegraphics[width=0.2\linewidth]{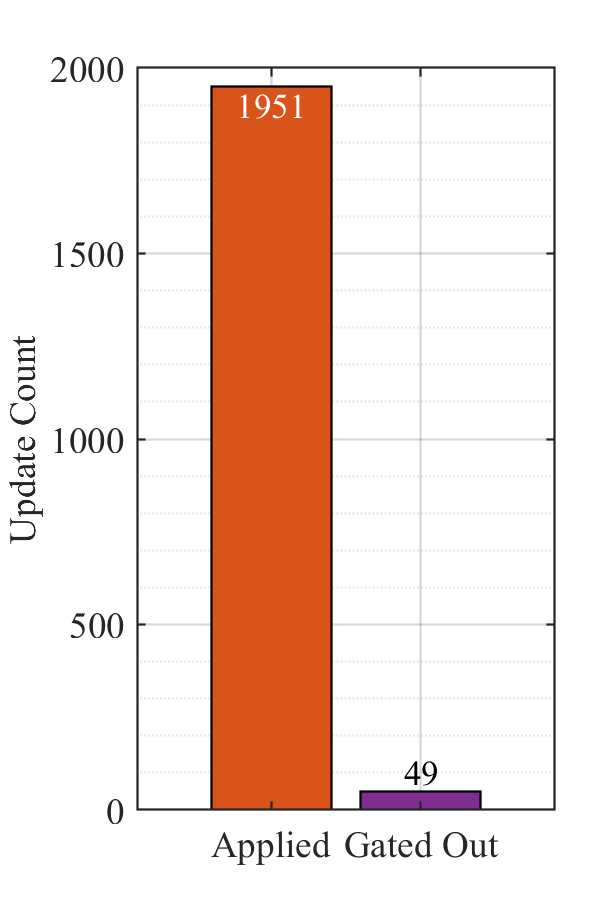}
    \caption{Number of ODM pseudo-measurement updates applied and rejected by the residual gate under nominal conditions.}
    \label{fig:favGateCount}
\end{figure}

The internal adaptive-fusion diagnostics are shown in \autoref{fig:favGnssScale}--\autoref{fig:favGateCount}. In \autoref{fig:favGnssScale}, throughout the processing interval, the GNSS stress remains tightly bounded, with a mean of $\mathbb{E}\!\left[s_{\mathrm{GNSS}}\right]=1.742$, well within the imposed upper limit $s_{\mathrm{GNSS},\max}=4.0$. This indicates that, although the adaptive fusion mechanism remains active, it correctly recognises the consistent reliability of the nominal simulated GNSS measurements and keeps the system in a relatively unstressed regime.

The state residual and adaptive covariance of the ODM pseudo-measurements are shown in \autoref{fig:favRsigma}, where the Euclidean magnitudes of the componentwise quantities are used for presentation and analysis. The residuals remain small, with $\mathbb{E}\!\left[\lVert e^r \rVert\right]=0.338$ m and $\mathbb{E}\!\left[\lVert e^v \rVert\right]=0.0951$ m/s. Combined with the low GNSS stress, this results in stable covariances, with $\mathbb{E}\!\left[\lVert \sigma^r \rVert\right]=0.191$ m and $\mathbb{E}\!\left[\lVert \sigma^v \rVert\right]=0.00632$ m/s, which are reduced relative to the baseline covariance at all epochs while remaining above the lower bounds. This demonstrates that, under nominal conditions, the ODM pseudo-measurement is assigned meaningful yet well-controlled confidence, as it is neither excessively inflated nor collapsed to its lower bound.

This interpretation is further supported by \autoref{fig:favGateCount}. Of the 2000 total pseudo-measurement updates, 1951 are applied and only 49 are rejected by the residual gate, corresponding to an acceptance rate of 97.55\%. The orbit aid is accepted at most update epochs because the ODM prediction is highly consistent with the VTL solution, a result attributable to the high reliability of the GNSS signal and the substantial influence of the initial states supplied to the ODM by the VTL solution.

Overall, the nominal-condition results demonstrate two key points. First, both VTL and OAVT provide a substantial improvement over STL in position and velocity estimation under high-quality spaceborne GPS measurements. Second, the proposed OAVT architecture maintains essentially the same accuracy as standalone VTL under benign tracking conditions, owing to the strong consistency between the ODM-derived information and the GNSS solution. This should be regarded as a desirable property rather than a limitation. The adaptive orbit aid does not compromise nominal performance and can therefore be interpreted as a robustness-enhancement layer whose principal benefit emerges under the degraded conditions analysed in the next subsection.

\subsection{Performance under degraded conditions}
\label{subsec:results_degraded}

\begin{figure}[htbp]
    \centering
    \includegraphics[width=0.5\linewidth]{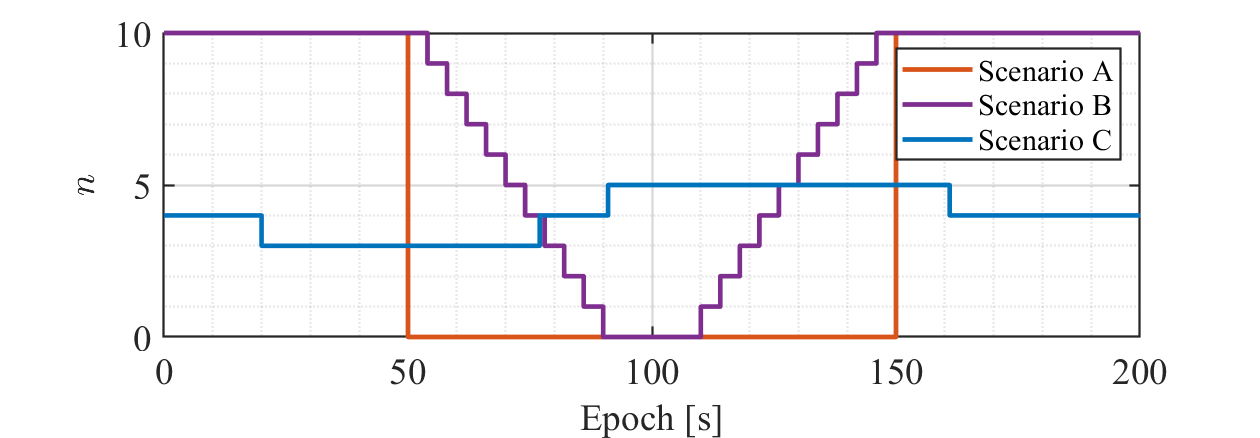}
    \caption{GNSS visibility profiles of Scenario A (full signal outage), B (gradual signal loss and reacquisition), and C (variable visibility).}
    \label{fig:configDegraded}
\end{figure}

The degraded-condition evaluation is performed using simulated GPS L1 IF data generated with the \texttt{gps-sdr-sim-master} simulator \citep{ebinuma2025gpssim}. In this case, the receiver is simulated as a LEO spaceborne receiver at an altitude of 422 km. This dataset is used to construct stressed tracking conditions that are directly relevant to the robustness objectives of the proposed algorithm, including limited GNSS satellite visibility and signal outages. Similar to the nominal evaluation in \autoref{subsec:results_nominal}, the data are processed over an interval of 200 s. The performance of the algorithm is evaluated across three constructed degraded-condition scenarios: 
\begin{enumerate}
    \item Scenario A simulates an abrupt, full signal outage lasting 100 s, resembling a receiver off-duty interval during a CubeSat duty cycle or a temporary hardware malfunction. This outage duration is chosen because, for a CubeSat with a typical duty cycle of 10 minutes \citep{narayana2022hummingbird,lantto2018poddutycycle}, it is reasonable to assume that the spaceborne receiver must remain on duty for 80\% to 90\% of the cycle to satisfy PNT service requirements, leaving an off-duty interval of 60 to 120 s that must be bridged using the ODM solution alone.
    \item Scenario B simulates a gradual loss and reacquisition of signal over 100 s, resembling an intermittent outage caused by geometric unavailability arising from satellite orbit and attitude dynamics, or changes in antenna operating status, with a total interval identical to that of Scenario A.
    \item Scenario C simulates frequent variations in GNSS satellite visibility, representing a common operational condition for spaceborne receivers. The visibility profile is derived from simulated GPS constellation access data relative to a GPS antenna onboard a LEO satellite in a 422 km polar orbit, from 25-06-2025 04:00:00 Coordinated Universal Time, Gregorian (UTCG) to 25-06-2025 04:10:00 UTCG.
\end{enumerate}
The specific GNSS visibility profiles of the three scenarios are illustrated in \autoref{fig:configDegraded}.

Across all degraded-condition scenarios, the number of visible GNSS satellites falls below four during certain intervals. Consequently, conventional least-squares (LS) STL navigation cannot produce a complete position--clock solution during these periods. In contrast, VTL and OAVT maintain continuous state estimates due to the presence of dynamic models; the relevant comparison is therefore their estimation accuracy, and the STL results are omitted from the following analysis.

\subsubsection{Scenario A: full signal outage}
\label{subsec:scenario_A}

\begin{figure}[htbp]
    \centering
    \includegraphics[width=0.9\linewidth]{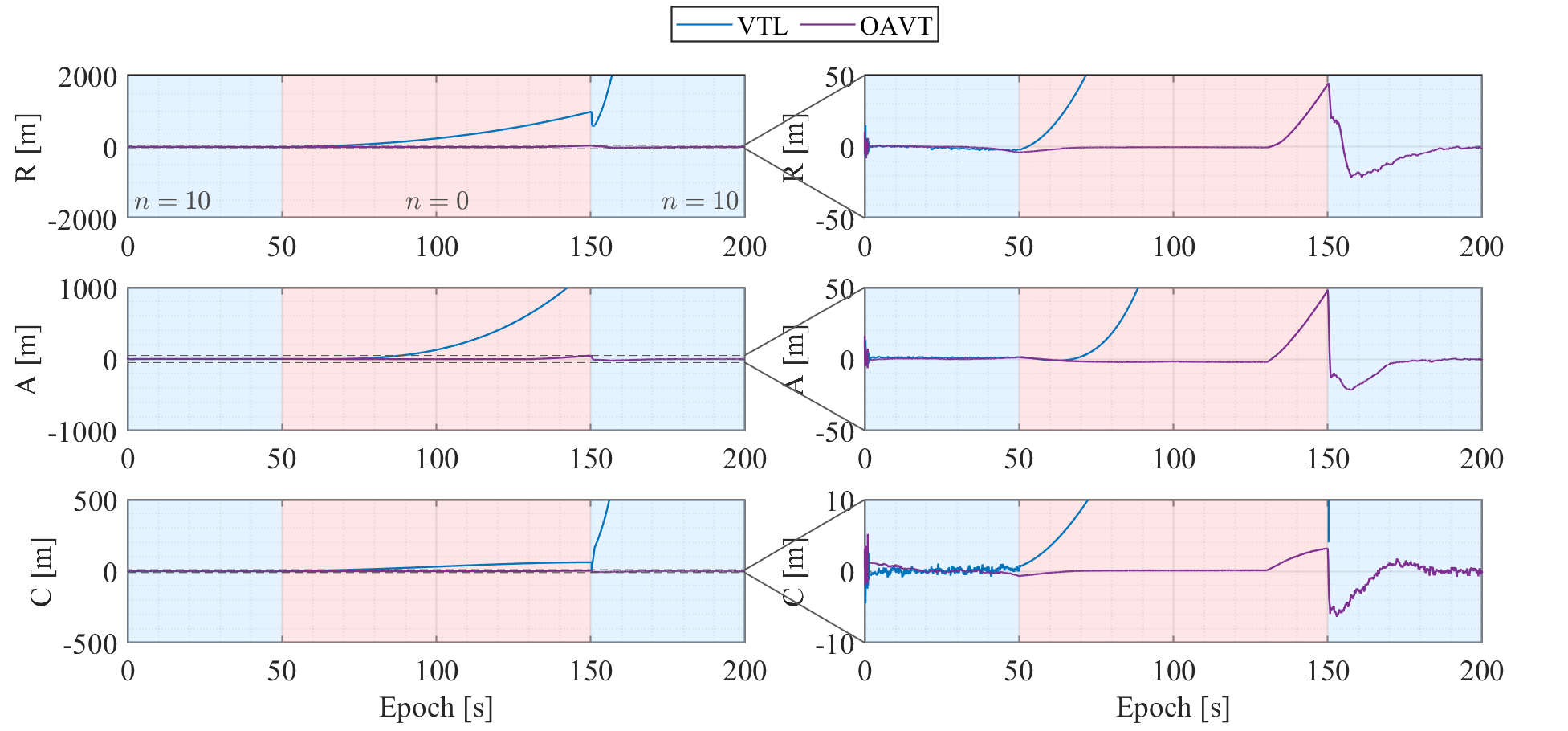}
    \caption{Position errors of VTL and OAVT in radial, along-track, and cross-track components under Scenario A.}
    \label{fig:S1PosRAC}
\end{figure}

\begin{figure}[htbp]
    \centering
    \includegraphics[width=0.9\linewidth]{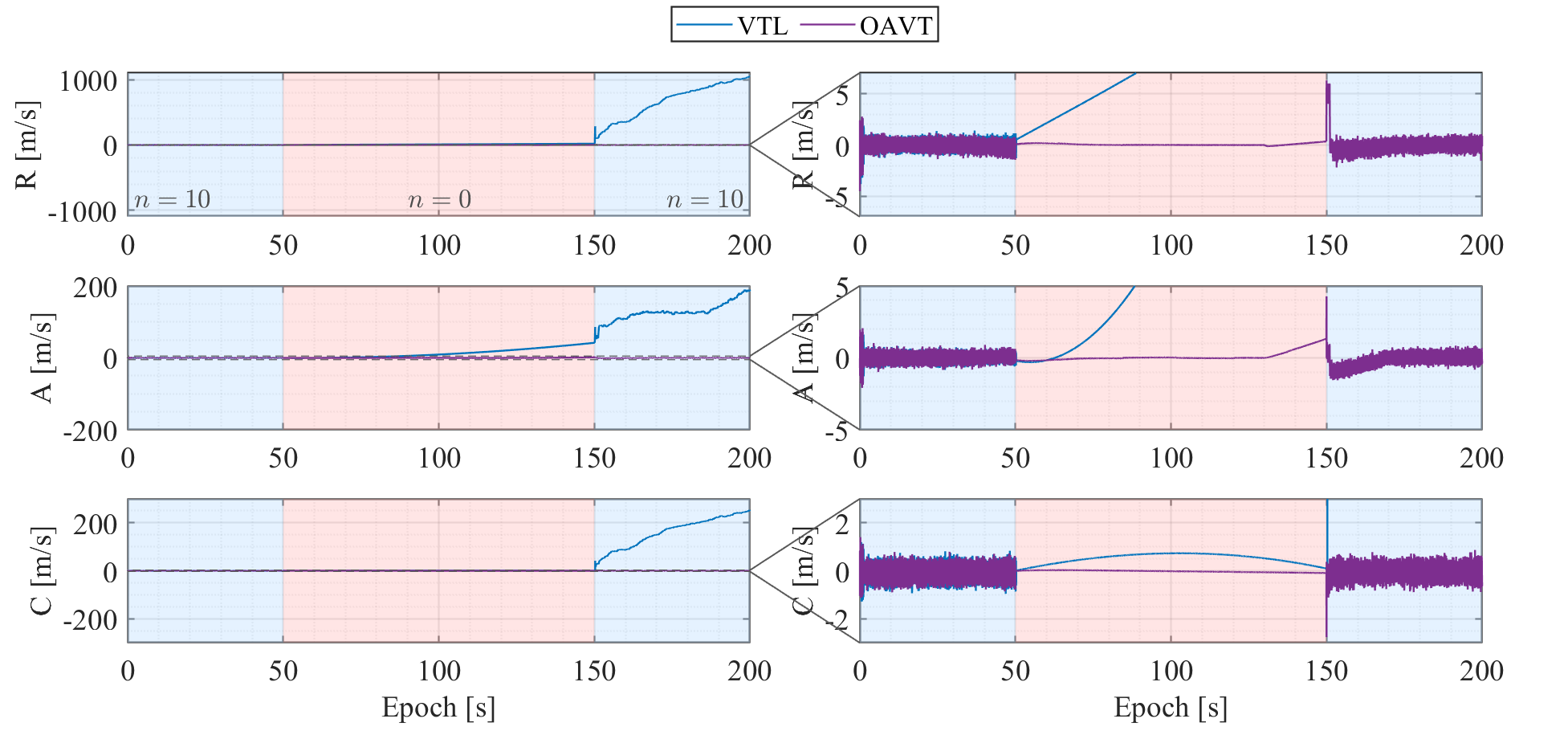}
    \caption{Velocity errors of VTL and OAVT in radial, along-track, and cross-track components under Scenario A.}
    \label{fig:S1VelRAC}
\end{figure}

\begin{figure}[htbp]
    \centering
    \includegraphics[width=0.9\linewidth]{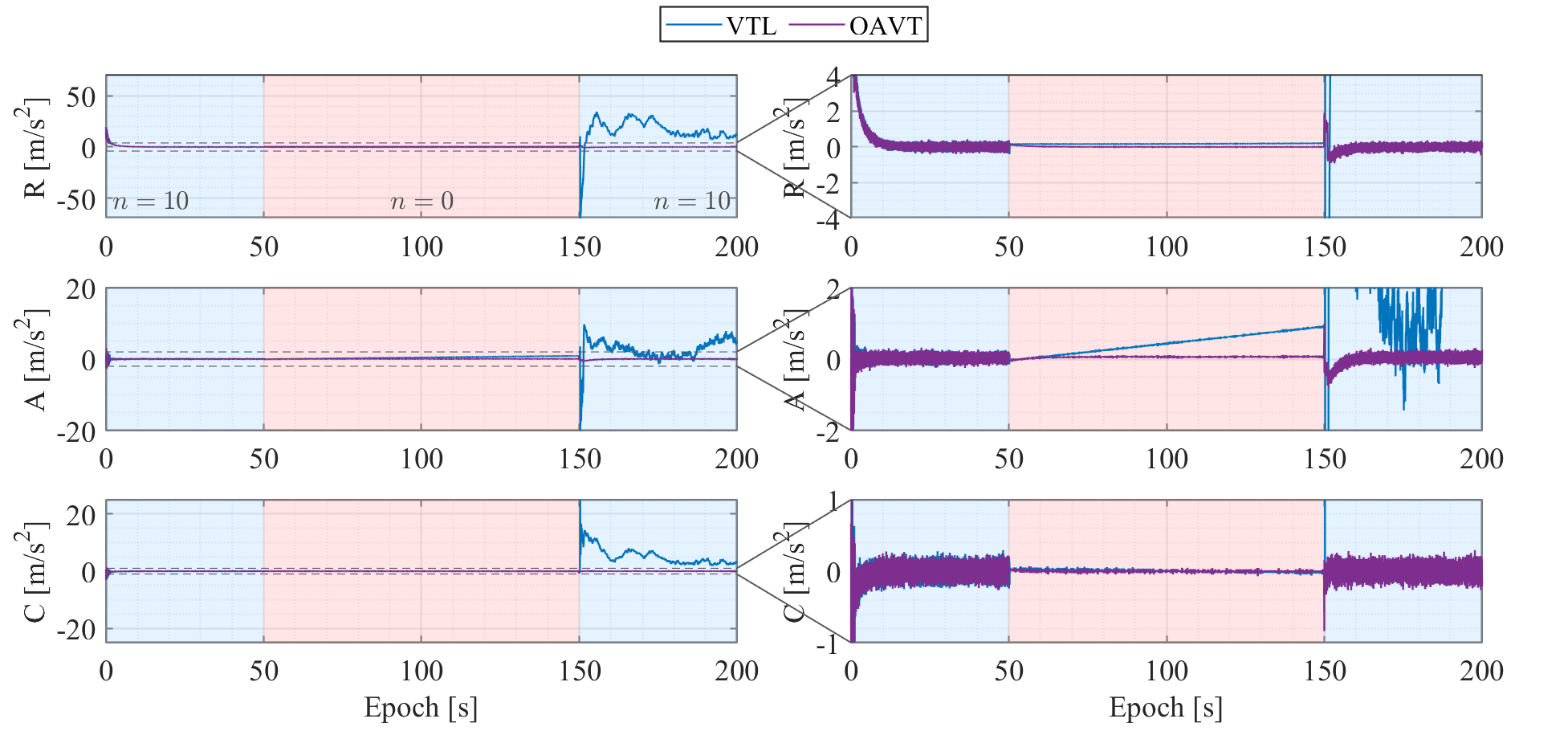}
    \caption{Acceleration errors of VTL and OAVT in radial, along-track, and cross-track components under Scenario A.}
    \label{fig:S1AccRAC}
\end{figure}

Scenario A imposes a complete GNSS outage from 50~s to 150~s. The RAC position, velocity, and acceleration (PVA) errors are presented in \autoref{fig:S1PosRAC}, \autoref{fig:S1VelRAC}, and \autoref{fig:S1AccRAC}, respectively; the right-hand panels provide magnified views of the low-error regime shown in the left-hand panels. The background shading denotes the variation in GNSS satellite visibility.

During the initial full-visibility interval, both VTL and OAVT achieve accurate estimates, as expected. Following the onset of the outage, the VTL position and velocity errors grow monotonically in all three RAC components because the orbit-related constraints are unavailable, whereas the acceleration errors remain comparatively small and slowly varying throughout the outage. When GNSS observations resume at 150~s, the line-of-sight parameters and code errors predicted from the diverged VTL solution can no longer support DLL re-tracking of the lost signals. Consequently, instead of returning to the correct course, the erroneous observables drive the solution away more rapidly during the remainder of the test.

In contrast, the OAVT solution remains bounded throughout the outage. During the outage, its position errors increase only gradually, while the velocity and acceleration errors remain close to zero in all three components. Upon measurement restoration, OAVT exhibits brief, finite transient errors and overshoots, attributable to the mismatch between the low-fidelity ODM prediction and the restored GNSS measurements. The estimates subsequently converge rapidly to a stable and accurate state, demonstrating robust recovery following measurement restoration.

\begin{figure}[htbp]
    \centering
    \includegraphics[width=0.6\linewidth]{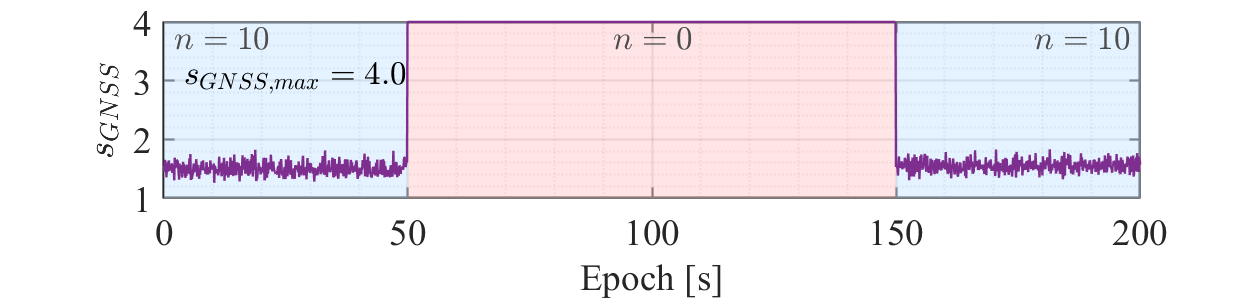}
    \caption{GNSS stress of OAVT under Scenario A.}
    \label{fig:S1GnssScale}
\end{figure}

\begin{figure}[htbp]
    \centering
    \includegraphics[width=0.5\linewidth]{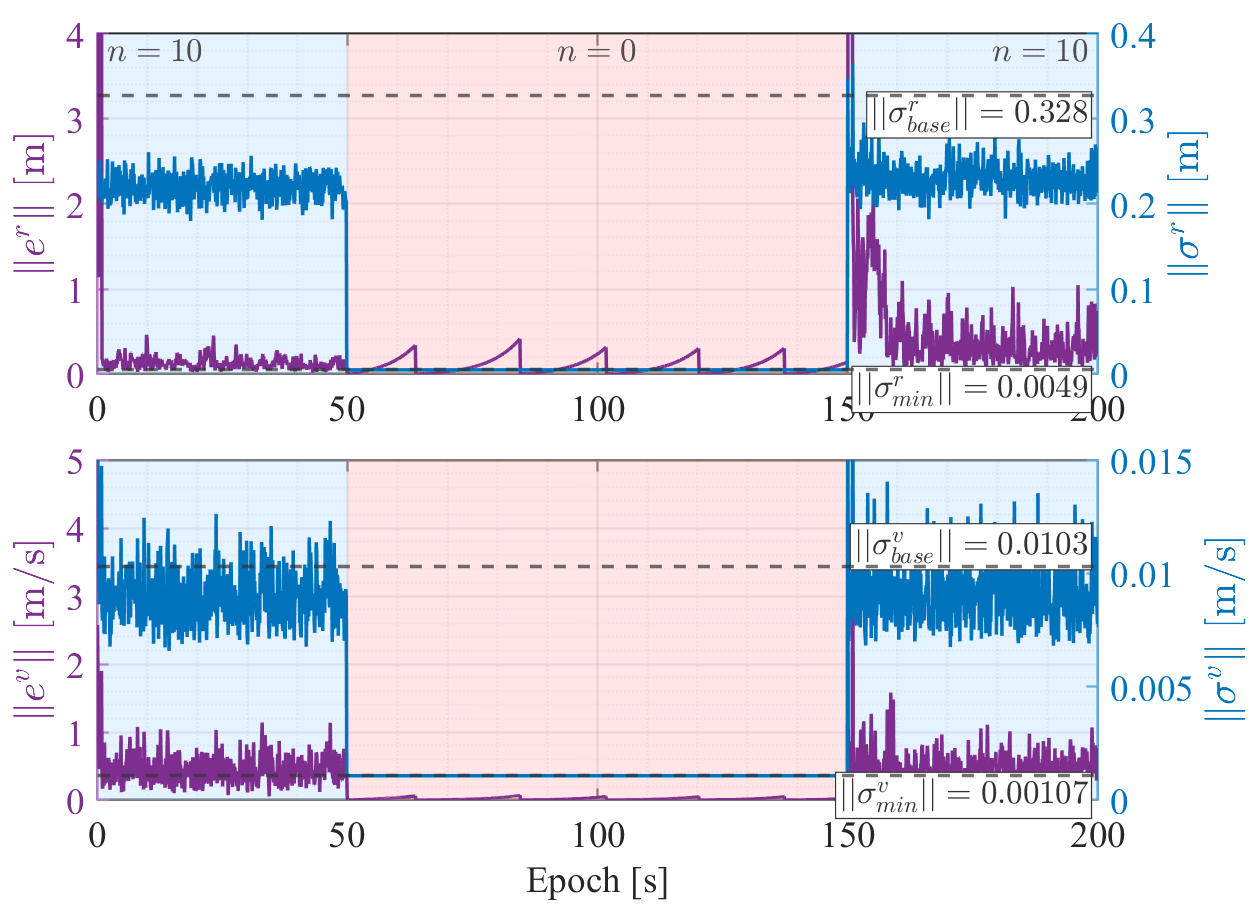}
    \caption{The state residual and adaptive covariance magnitudes of the ODM pseudo-measurements under Scenario A.}
    \label{fig:S1Rsigma}
\end{figure}

The internal diagnostics in \autoref{fig:S1GnssScale} and \autoref{fig:S1Rsigma} support this interpretation. At the onset of the outage, the GNSS stress immediately reaches its saturation limit, $s_{\mathrm{GNSS},\max}=4.0$, and remains saturated throughout the outage. The resulting covariance adaptation drives the ODM covariance towards its lower bound, indicating that the adaptive mechanism identifies the GNSS measurements as invalid and transfers estimation authority to the ODM. After GNSS measurement restoration at 150~s, the GNSS stress rapidly returns to its pre-outage level, while the adaptive covariance magnitudes recover to values close to their nominal baselines. At the restoration epoch, both the position and velocity residual norms exhibit pronounced spikes because of the discrepancy between the propagated ODM solution and the restored GNSS measurements, but these transients decay quickly thereafter. This behaviour is consistent with the brief overshoots observed in the PVA errors and confirms that the estimator re-establishes balanced ODM-GNSS fusion once reliable measurements become available again.

\begin{figure}[htbp]
    \centering
    \includegraphics[width=0.5\linewidth]{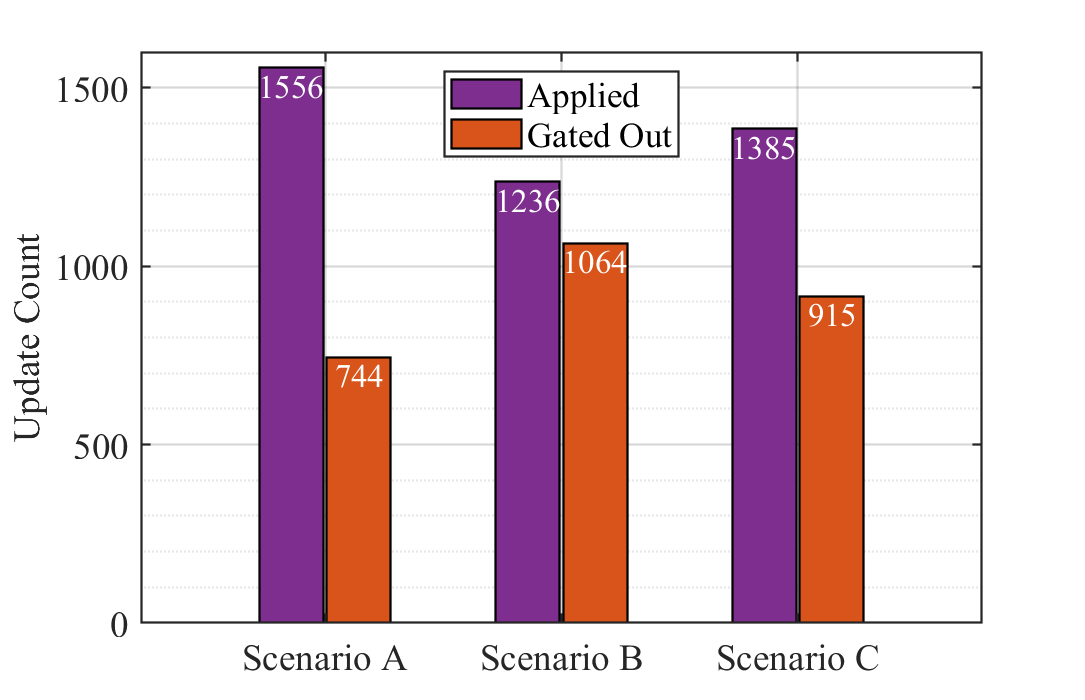}
    \caption{Number of ODM pseudo-measurement updates applied and rejected by the residual gate under Scenario A, B, and C.}
    \label{fig:degGateCount}
\end{figure}

The ODM pseudo-measurement update statistics in \autoref{fig:degGateCount} show that 1556 updates are accepted and 744 are rejected. These results confirm that residual gating remains active throughout the processing and prevents indiscriminate reliance on ODM pseudo-measurements. For both VTL and OAVT, the radial PVA errors are generally larger than their along-track and cross-track counterparts, consistent with the observations discussed in \autoref{subsec:results_nominal}. The PVA RMSEs of OAVT under Scenario A are detailed in \autoref{tab:degraded_rmse}.

\subsubsection{Scenario B: gradual signal loss and reacquisition}
\label{subsec:scenario_B}

\begin{figure}[htbp]
    \centering
    \includegraphics[width=0.9\linewidth]{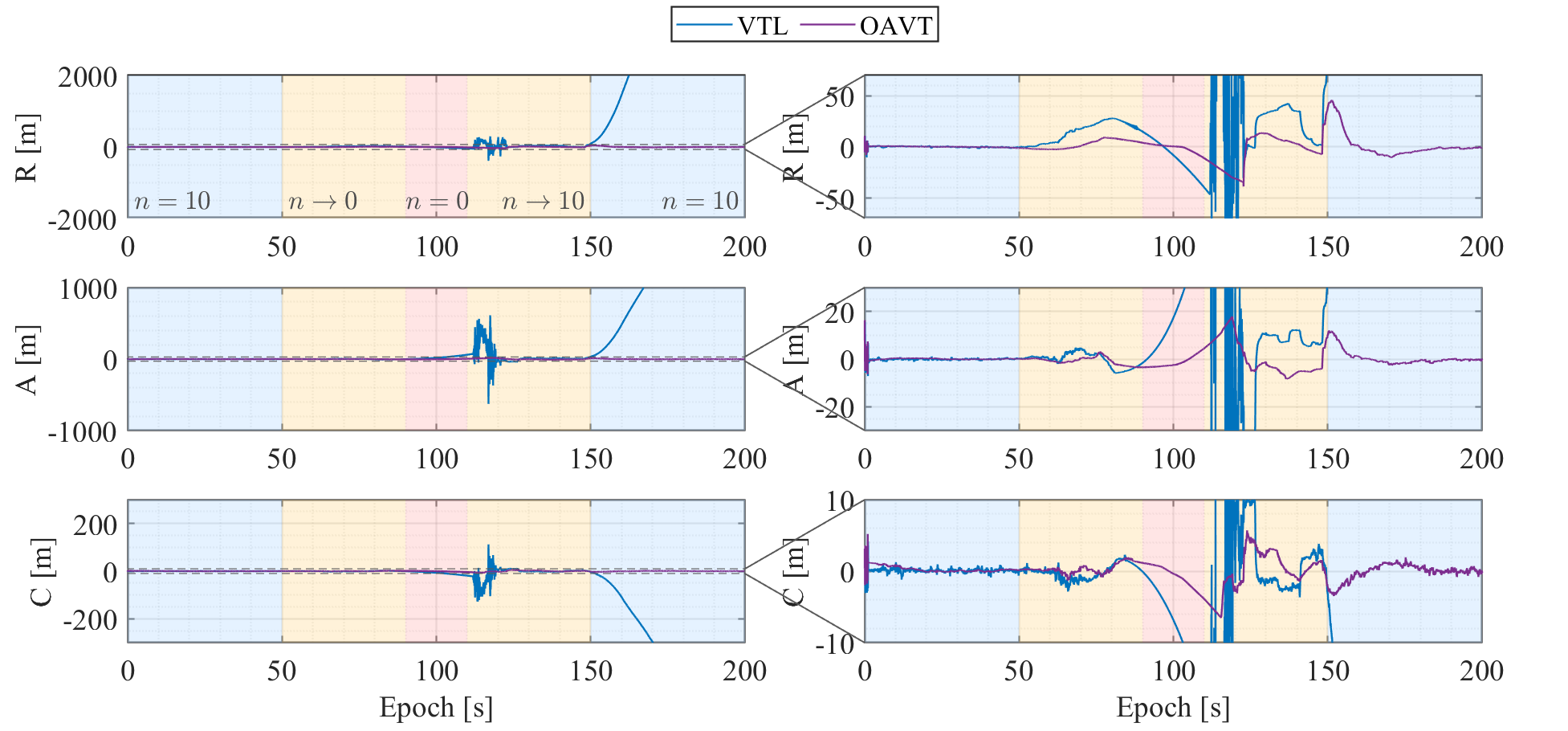}
    \caption{Position errors of VTL and OAVT in radial, along-track, and cross-track components under Scenario B.}
    \label{fig:S2PosRAC}
\end{figure}

Scenario B considers the progressive loss and subsequent reacquisition of GNSS signals. Satellite visibility decreases stepwise from \(n=10\) at 50~s to \(n=0\) at 90~s, remains at zero until 110~s, and subsequently recovers to \(n=10\) by 150~s.

\begin{figure}[htbp]
    \centering
    \includegraphics[width=0.9\linewidth]{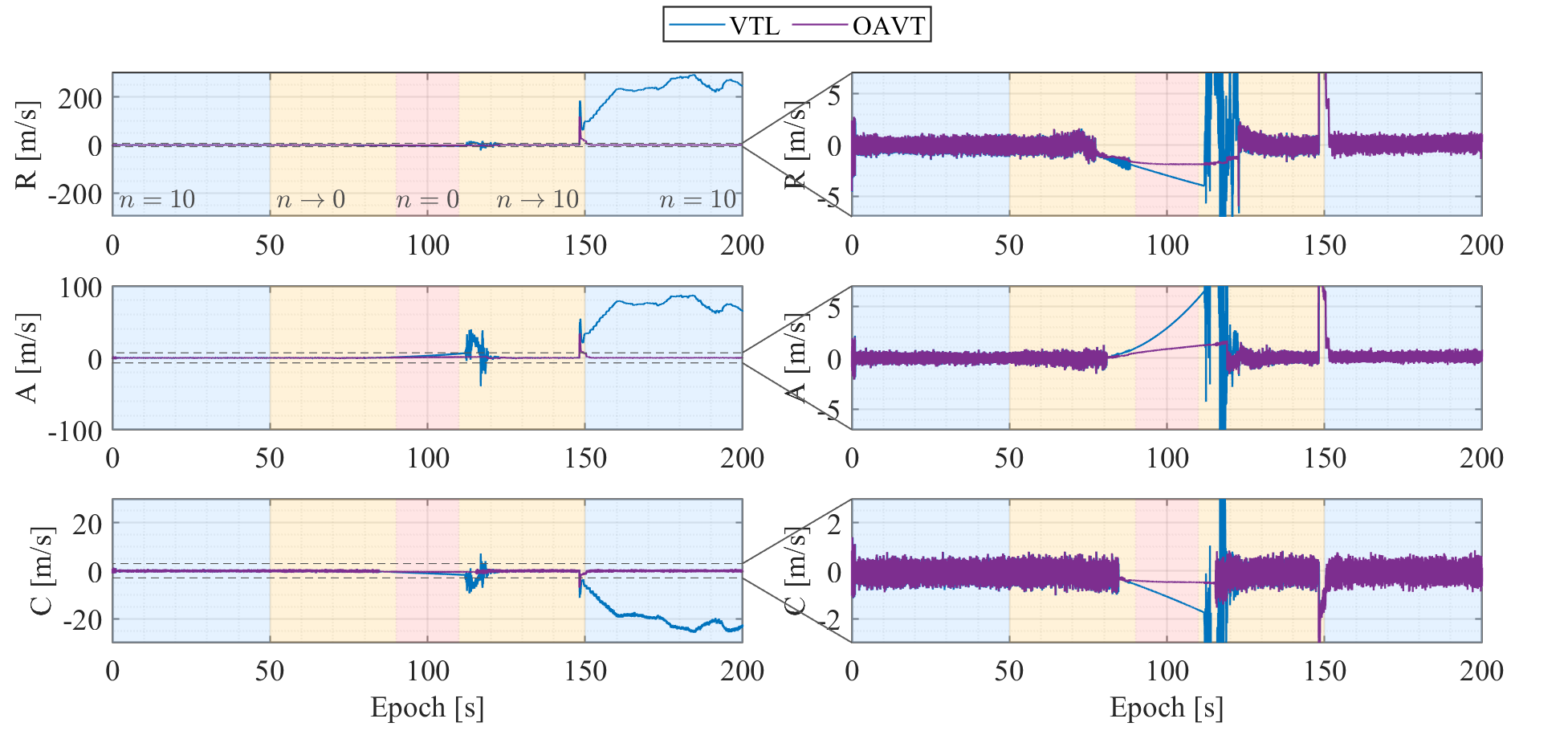}
    \caption{Velocity errors of VTL and OAVT in radial, along-track, and cross-track components under Scenario B.}
    \label{fig:S2VelRAC}
\end{figure}

\begin{figure}[htbp]
    \centering
    \includegraphics[width=0.9\linewidth]{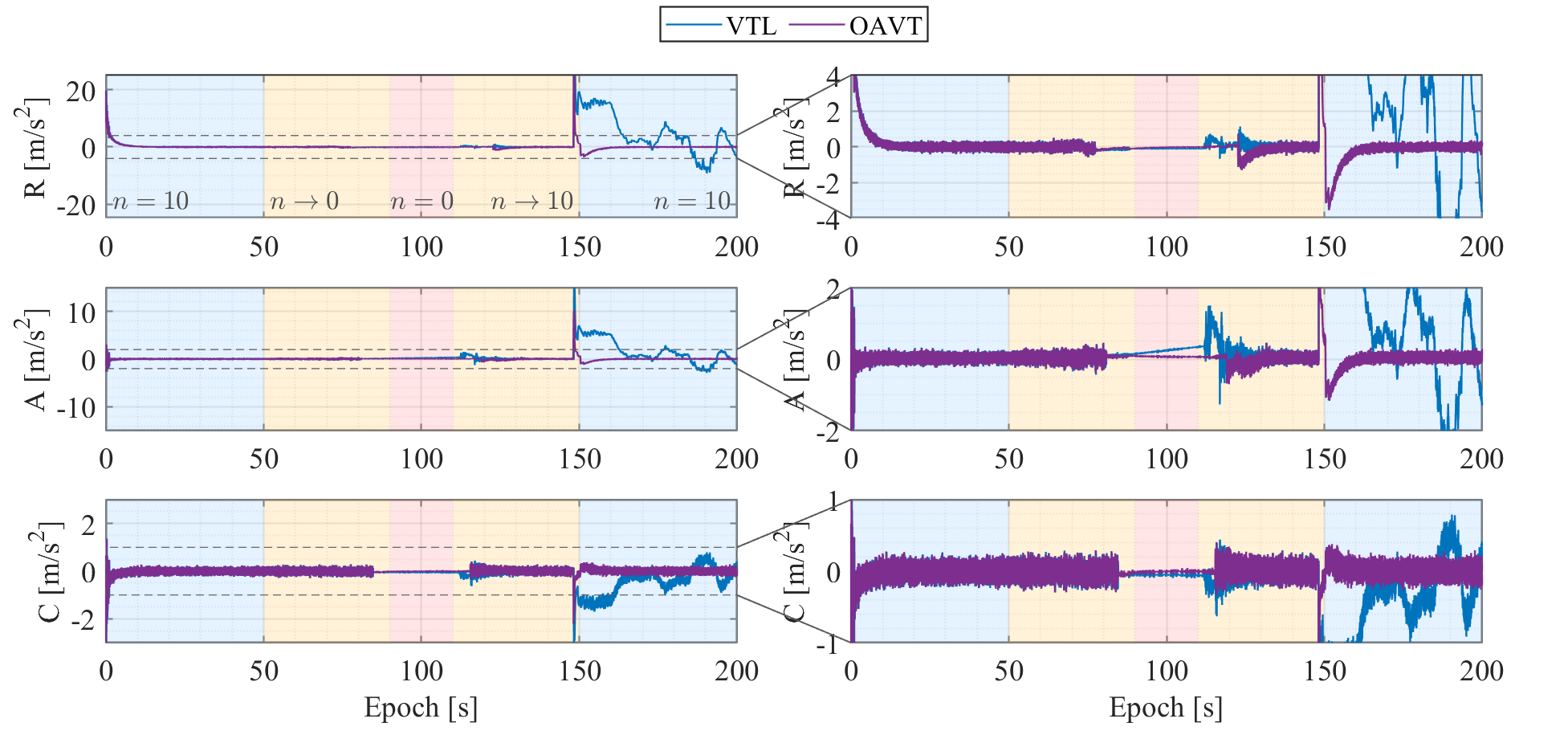}
    \caption{Acceleration errors of VTL and OAVT in radial, along-track, and cross-track components under Scenario B.}
    \label{fig:S2AccRAC}
\end{figure}

As shown in \autoref{fig:S2PosRAC}, \autoref{fig:S2VelRAC}, and \autoref{fig:S2AccRAC}, the VTL errors begin to increase as soon as satellite visibility decreases, even before the zero-visibility interval is reached. This behaviour demonstrates the susceptibility of standalone VTL to the cross-channel contamination discussed in \autoref{sec:intro}. As the number of tracked channels decreases, the navigation update is formed from increasingly stressed measurements under weak satellite geometry. The resulting estimation errors are then fed back to the remaining channels, thereby propagating and amplifying the degradation throughout the vector-tracking architecture. This effect is particularly evident during the early reacquisition interval between approximately 112~s and 123~s, when a partially restored but still insufficient measurement set produces notable oscillatory excursions in the position, velocity, and acceleration estimates. Although the tracking loops temporarily stabilise until approximately 148~s, the inconsistency between the restored measurements and the predicted states subsequently causes the standalone VTL to lose lock.

In contrast, although OAVT accuracy is affected by the high-GDOP measurements and the limited fidelity of the ODM, its state estimates remain tightly bounded throughout the degraded interval, with position errors remaining below approximately 50~m in the worst case. As visibility decreases and later recovers, the position errors exhibit gradual component-dependent drift, while the velocity and acceleration errors remain comparatively small. A brief transient appears near the final restoration epoch at approximately 150~s,  but it exerts negligible influence on the subsequent position estimates after the rapid correction.

\begin{figure}[htbp]
    \centering
    \includegraphics[width=0.6\linewidth]{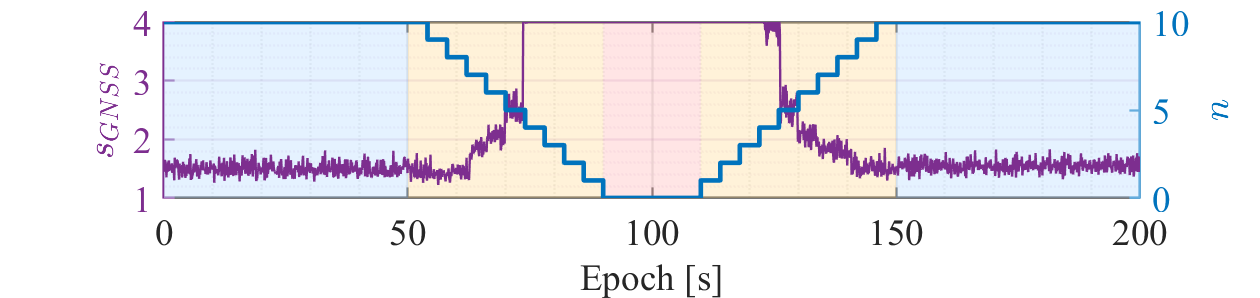}
    \caption{GNSS stress of OAVT and GNSS visibility profile of Scenario B.}
    \label{fig:S2GnssScale}
\end{figure}

\begin{figure}[htbp]
    \centering
    \includegraphics[width=0.5\linewidth]{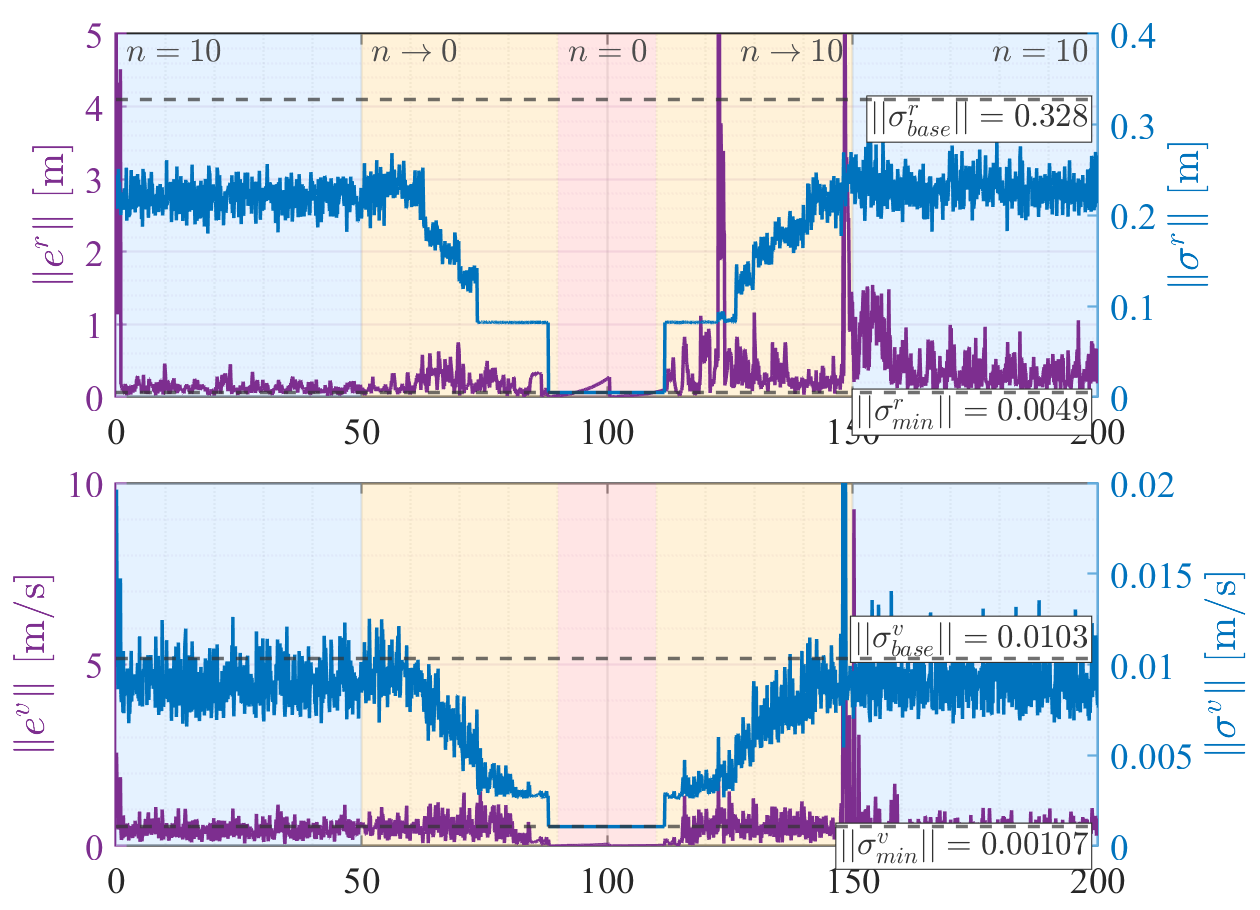}
    \caption{The state residual and adaptive covariance magnitudes of the ODM pseudo-measurements under Scenario B.}
    \label{fig:S2Rsigma}
\end{figure}

The GNSS stress shown in \autoref{fig:S2GnssScale} responds to the variation in measurement availability as intended. As satellite visibility decreases, \(s_{\mathrm{GNSS}}\) rises from its nominal level and reaches its upper limit during the interval of complete signal loss. It subsequently decreases during reacquisition, while remaining elevated during the intermediate low-visibility stages in which the restored measurements are still weak and only partially consistent with the propagated state. The GNSS-stress-dependent adaptive covariances in \autoref{fig:S2Rsigma} exhibit the reciprocal response, contracting towards their lower bounds during measurement loss and progressively expanding towards their nominal baselines as visibility and measurement consistency recover.

Compared with Scenario A, Scenario B produces more frequent ODM--GNSS residual excursions, particularly during the partial-visibility and final restoration intervals shown in \autoref{fig:S2Rsigma}. The residual gate comes into effect by preventing these abrupt corrections from destabilising the tracking loops. Its increased activity is reflected by the 46.3\% rejection rate reported in \autoref{fig:degGateCount}. The PVA RMSEs of OAVT under Scenario B are detailed in \autoref{tab:degraded_rmse}.

\subsubsection{Scenario C: variable visibility}
\label{subsec:scenario_C}

Scenario C represents sustained operation under marginal satellite availability and time-varying measurement geometry. Its visibility profile is derived from simulated GPS access data and varies between \(n=3\) and \(n=5\) without a complete outage. Visibility decreases from \(n=4\) to \(n=3\) at 20~s, increases successively to \(n=4\) and \(n=5\) between 78~s and 90~s, and returns to \(n=4\) at 161~s.

\begin{figure}[htbp]
    \centering
    \includegraphics[width=0.9\linewidth]{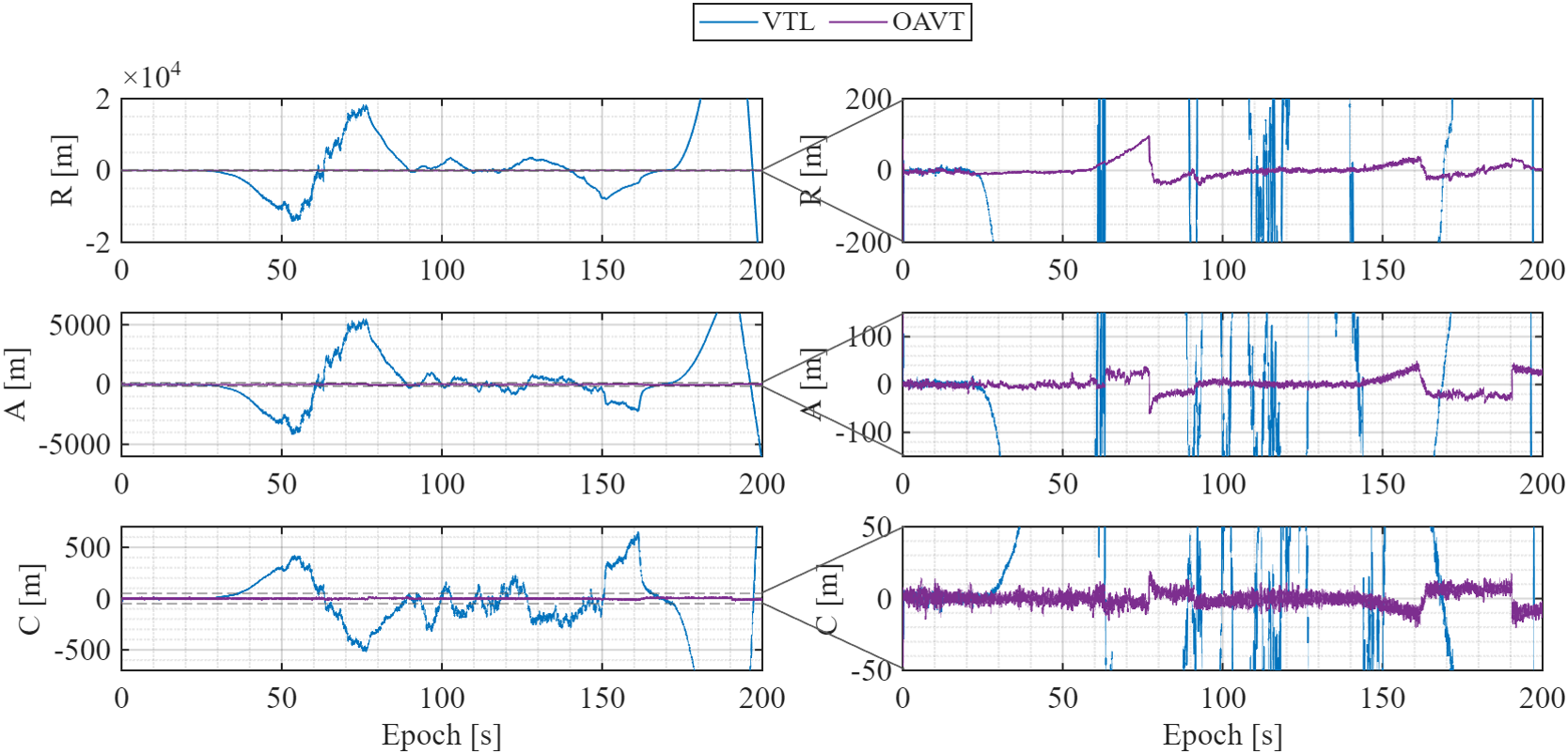}
    \caption{Position errors of VTL and OAVT in radial, along-track, and cross-track components under Scenario C.}
    \label{fig:S3PosRAC}
\end{figure}

\begin{figure}[htbp]
    \centering
    \includegraphics[width=0.9\linewidth]{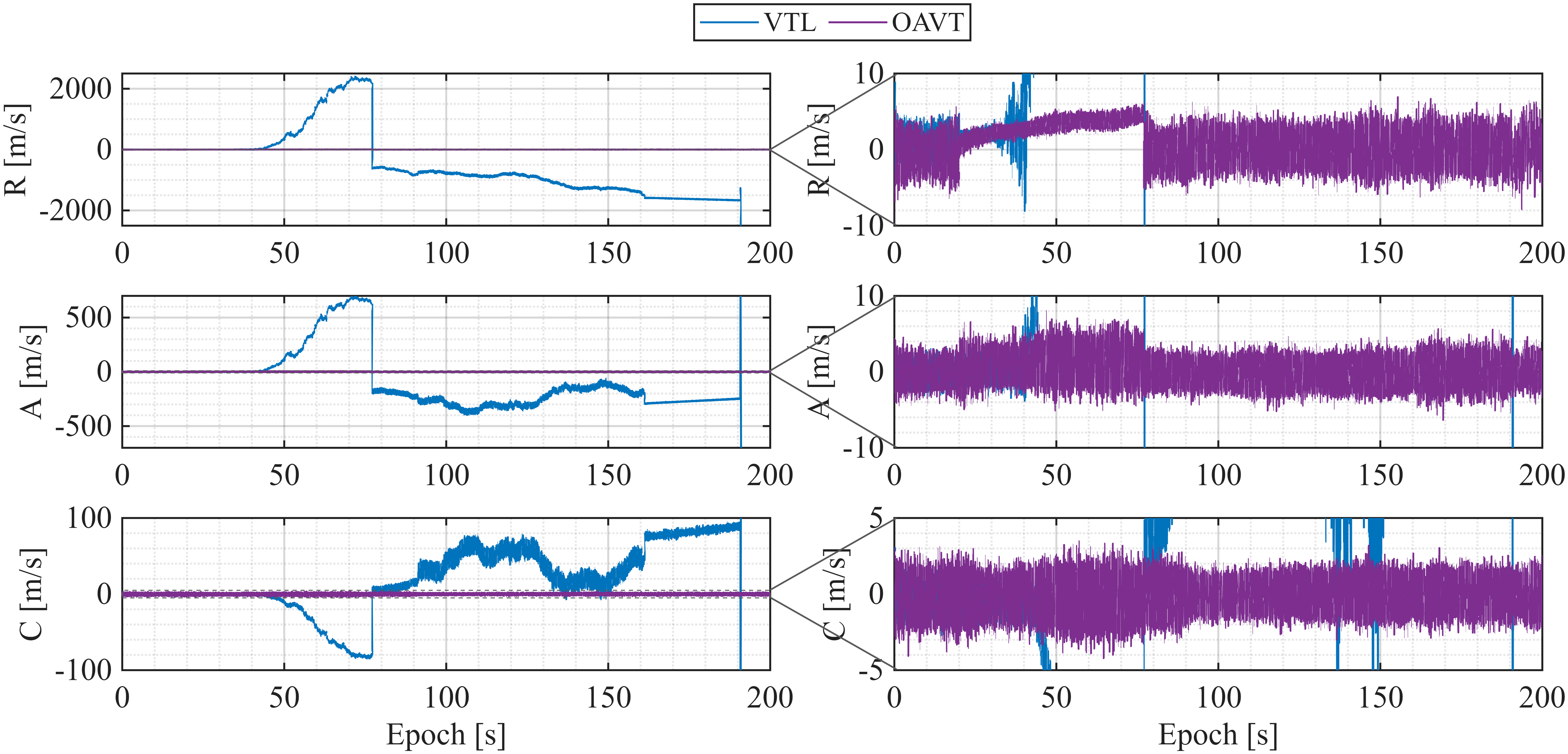}
    \caption{Velocity errors of VTL and OAVT in radial, along-track, and cross-track components under Scenario C.}
    \label{fig:S3VelRAC}
\end{figure}

\begin{figure}[htbp]
    \centering
    \includegraphics[width=0.9\linewidth]{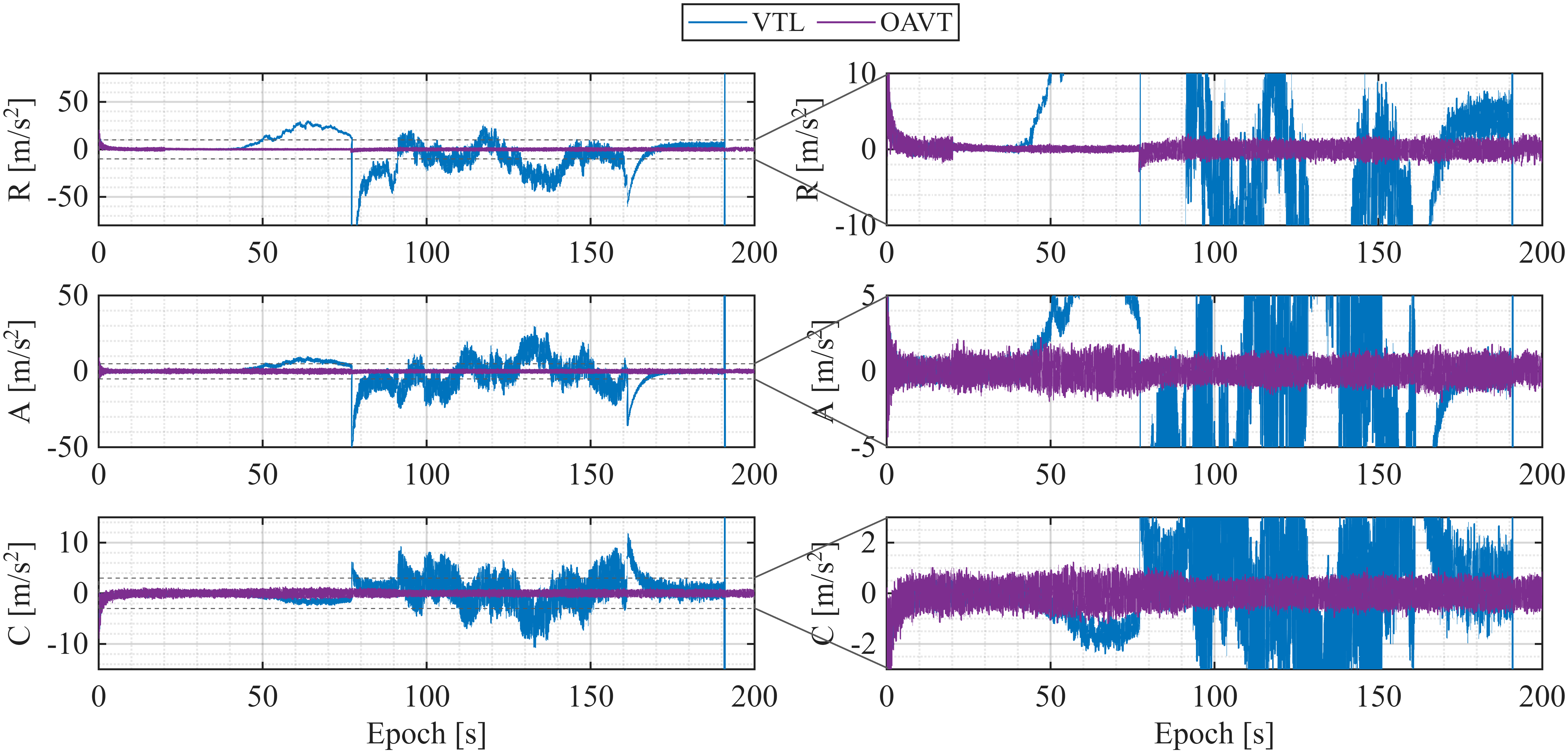}
    \caption{Acceleration errors of VTL and OAVT in radial, along-track, and cross-track components under Scenario C.}
    \label{fig:S3AccRAC}
\end{figure}

As shown in \autoref{fig:S3PosRAC}, \autoref{fig:S3VelRAC}, and \autoref{fig:S3AccRAC}, the standalone VTL becomes unstable shortly after visibility decreases to \(n=3\). Although the filter maintains state continuity through model propagation, the available GNSS measurements are insufficient to form a complete position--clock update and provide only weak geometric constraints. Incorporating these low-quality measurements progressively contaminates the navigation estimate, whose errors are subsequently fed back to the remaining tracking channels. The resulting cross-channel error propagation produces very large excursions in all RAC components, with errors several orders of magnitude greater than those of OAVT. Although visibility subsequently improves to \(n\geq4\), the accumulated state errors prevent the VTL from recovering consistently, eventually resulting in a complete loss of lock at approximately 190~s.

In contrast, OAVT achieves substantially better performance across all PVA states throughout the test. Its accuracy is nevertheless affected by the prolonged \(n=3\) interval and persistently weak measurement geometry, as evidenced by the larger errors relative to Scenarios A and B. The observed fluctuations arise from measurement noise, changing geometry, and transient inconsistencies between the ODM predictions and GNSS estimates. Despite these adverse conditions, OAVT maintains robust tracking, limiting the maximum position and velocity errors in one direction to approximately 100~m and 6~m/s, respectively. The largest position deviations occur near changes in satellite visibility. During the \(n=3\) interval, the system assigns greater authority to the ODM, resulting in a period of bounded but finite estimation drift. When visibility increases to \(n=4\) at 78~s, allowing the GNSS filter to form a complete solution, the fusion algorithm corrects the accumulated error within a single update epoch without destabilising the tracking loops. These results demonstrate that OAVT preserves an adaptive and robust navigation solution under realistic visibility variations for which neither STL nor VTL can guarantee solution continuity.

\begin{figure}[htbp]
    \centering
    \includegraphics[width=0.6\linewidth]{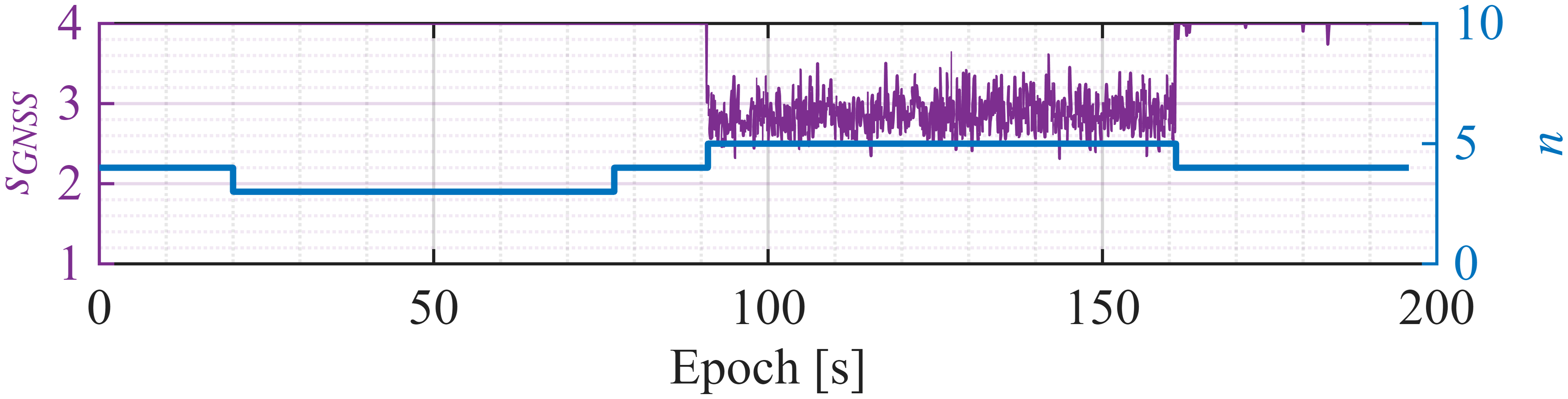}
    \caption{GNSS stress of OAVT and GNSS visibility profile of Scenario C.}
    \label{fig:S3GnssScale}
\end{figure}

The GNSS stress shown in \autoref{fig:S3GnssScale} reflects the limited information and poor geometry associated with the available GNSS measurements. It remains at its upper limit, \(s_{\mathrm{GNSS},\max}=4.0\), throughout the \(n=3\) and \(n=4\) intervals. Only when visibility increases to \(n=5\) at 90~s does \(s_{\mathrm{GNSS}}\) decrease to an intermediate mean level of approximately 2.5, thereby restoring greater authority to the GNSS observations. Nevertheless, this value remains substantially higher than the nominal mean stress of approximately 0.6 observed in the other scenarios.

\begin{figure}[htbp]
    \centering
    \includegraphics[width=0.5\linewidth]{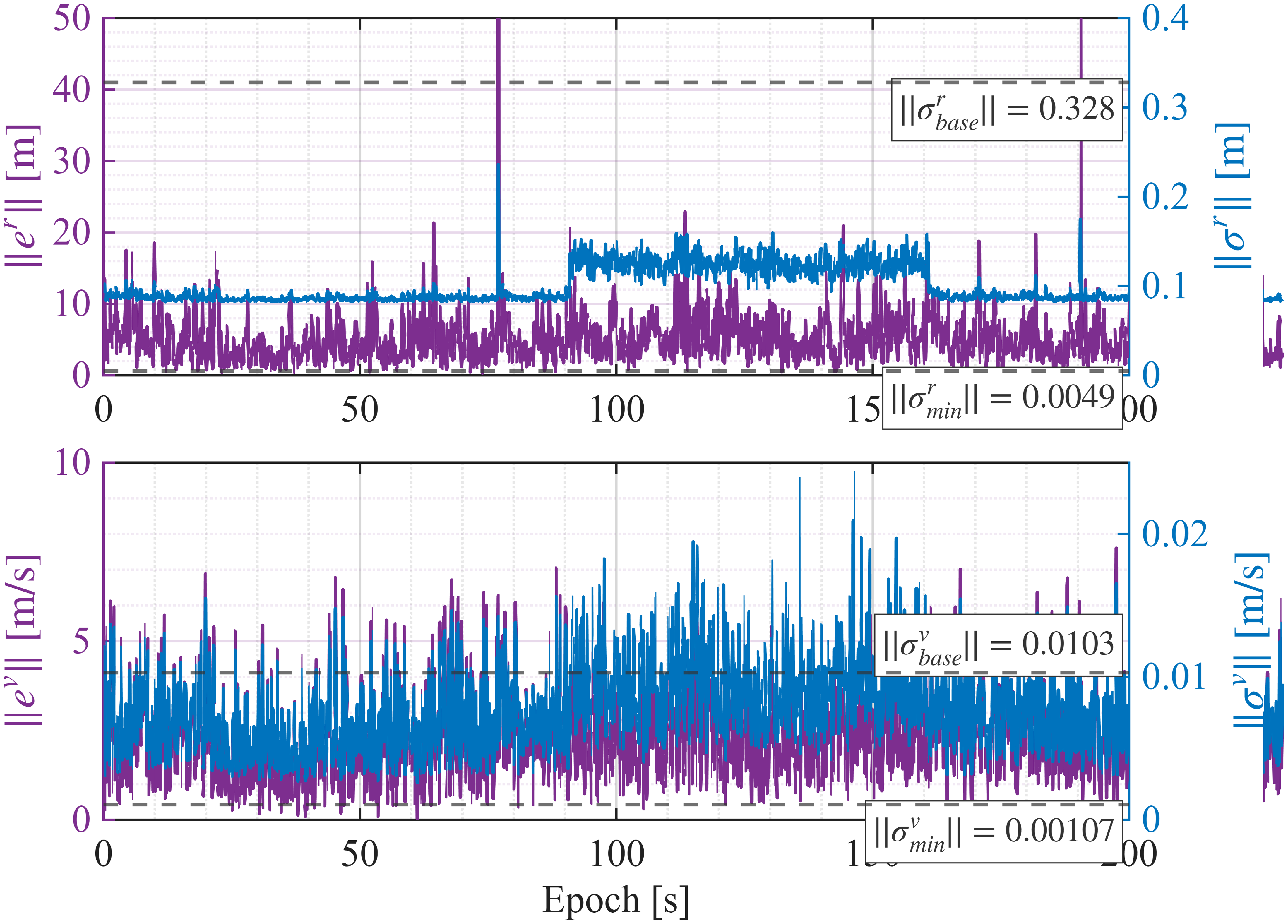}
    \caption{The state residual and adaptive covariance magnitudes of the ODM pseudo-measurements under Scenario C.}
    \label{fig:S3Rsigma}
\end{figure}

The adaptive covariance histories in \autoref{fig:S3Rsigma} exhibit the corresponding reciprocal behaviour. While the GNSS stress is saturated, the ODM covariance contracts to increase the relative contribution of the ODM pseudo-measurements. When visibility increases to \(n=5\) and the GNSS stress decreases, the adaptive covariance expands accordingly. Because the ODM--GNSS residuals remain persistently elevated and fluctuate frequently, the covariance does not collapse to its lower bound, as it does when GNSS measurements are completely absent. Retaining this covariance margin proves critical at visibility-transition epochs, as it allows accumulated ODM-induced deviations to be corrected when sufficient GNSS information becomes available.

As reported in \autoref{fig:degGateCount}, the residual gate rejects 39.8\% of the ODM pseudo-measurement updates. This substantial rejection rate confirms that the gate actively prevents abrupt and potentially destabilising corrections derived from excessively inconsistent pseudo-measurements. The combined covariance adaptation and residual gating therefore enable OAVT to exploit the continuity and short-term accuracy provided by the ODM while limiting its adverse influence on GNSS signal tracking. The PVA RMSEs of OAVT under Scenario C are detailed in \autoref{tab:degraded_rmse}.

\begin{table}[htbp]
\caption{OAVT RMSE under degraded conditions.}
\label{tab:degraded_rmse}
\centering
\small
\setlength{\tabcolsep}{4pt}
\renewcommand{\arraystretch}{1.1}
\begin{tabular}{|>{\centering\arraybackslash}p{0.17\linewidth}|>{\centering\arraybackslash}p{0.12\linewidth}|>{\centering\arraybackslash}p{0.18\linewidth}|>{\centering\arraybackslash}p{0.18\linewidth}|>{\centering\arraybackslash}p{0.18\linewidth}|}
\hline
\textbf{Quantity} & \textbf{Component} & \textbf{Scenario A} & \textbf{Scenario B} & \textbf{Scenario C} \\
\hline
\multirow{4}{*}{Position error [m]}
 & Radial      & 5.37 & 9.42 & 20.45 \\
\cline{2-5}
 & Along-track & 6.21 & 3.55 & 16.26 \\
\cline{2-5}
 & Cross-track & 0.94 & 1.44 & 5.27 \\
\cline{2-5}
 & 3D          & 8.27 & 10.17 & 26.65 \\
\hline
\multirow{4}{*}{Velocity error [m/s]}
 & Radial      & 0.43 & 4.37 & 2.12 \\
\cline{2-5}
 & Along-track & 0.27 & 1.39 & 1.36 \\
\cline{2-5}
 & Cross-track & 0.17 & 0.36 & 0.27 \\
\cline{2-5}
 & 3D          & 0.53 & 4.60 & 2.64 \\
\hline
\multirow{4}{*}{Acceleration error [m/s$^2$]}
 & Radial      & 0.52 & 1.41 & 0.69 \\
\cline{2-5}
 & Along-track & 0.10 & 0.39 & 0.41 \\
\cline{2-5}
 & Cross-track & 0.08 & 0.11 & 0.36 \\
\cline{2-5}
 & 3D          & 0.53 & 1.47 & 0.88 \\
\hline
\end{tabular}
\end{table}

\section{Discussion}
\label{sec:discussion}

\subsection{Computational complexity and onboard feasibility}
\label{subsec:discussion_complexity}
The proposed OAVT preserves the dominant computational structure of the baseline VTL. At GNSS epoch $k$, let $n_k$ denote the number of active channels, let $m=11$ denote the navigation-state dimension, and let $N_{\mathrm{samp}}$ denote the number of front-end samples processed per channel over one PDI. The per-channel correlation, carrier mixing, and integrate-and-dump operations are linear in $N_{\mathrm{samp}}$, so the front-end and measurement-formation cost scales as $\mathcal{O}(n_k N_{\mathrm{samp}})$. After this step, the EKF update forms and inverts the innovation covariance with dimension $2n_k \times 2n_k$, giving the baseline VTL complexity
\begin{equation}
    C_{\mathrm{VTL}}(k)=\mathcal{O}\!\left(n_k N_{\mathrm{samp}}\right)+\mathcal{O}\!\left(m n_k^2\right)+\mathcal{O}\!\left(n_k^3\right).
\end{equation}
    Since $m=11$ is fixed and small, this may be written more compactly as
\begin{equation}
    C_{\mathrm{VTL}}(k)=\mathcal{O}\!\left(n_k N_{\mathrm{samp}}+n_k^3\right).
\end{equation}

OAVT retains this VTL pipeline and adds only a reduced-rate ODM branch. At each scheduled ODM update, the pseudo-measurement fusion operates on a fixed six-dimensional position-velocity pseudo-measurement, so its matrix operations are $\mathcal{O}(1)$ with respect to $n_k$. Let $N_{\mathrm{ODM}}$ denote the number of GNSS epochs between two consecutive ODM updates, and let $L$ denote the number of internal numerical integration steps required by the propagator over one ODM horizon. The amortised OAVT complexity per GNSS epoch is then
\begin{equation}
    C_{\mathrm{OAVT}}(k)=\mathcal{O}\!\left(n_k N_{\mathrm{samp}}+n_k^3+\frac{L}{N_{\mathrm{ODM}}}\right),
\end{equation}
where the constant-cost pseudo-measurement fusion has been absorbed into the hidden factor. Therefore, OAVT does not alter the channel-dependent scaling of VTL; it only adds the amortised cost of a reduced-rate orbit propagation. This is the key practical result: the substantial performance gains observed in \autoref{subsec:results_degraded} are obtained without introducing any additional per-channel or per-sample processing loop. For CubeSat orbit determination, where onboard resources are limited, the added burden is therefore moderate and remains compatible with the intended low-SWaP application.

\subsection{Algorithm parameter tuning}
\label{subsec:discussion_tuning}

The pseudo-measurement fusion parameters, including the ODM update interval, the sensitivity gains, the reference scales, the covariance bounds, and the residual gate thresholds, are tuned empirically through iterative optimisation. The objective of this process is not to achieve the best performance for a particular test dataset, but to obtain a parameter set that is robust and adaptive across changing observation conditions. In this sense, parameter tuning is treated as a robustness-design problem rather than a single-point performance optimisation problem.

From the tuning iterations, a practical guideline is that the ODM should be updated at least every 200 GNSS filter epochs, i.e. $T_{\mathrm{ODM}} \leq 200T$. If $T_{\mathrm{ODM}}$ is chosen larger than this, its contribution can become too weak to make a meaningful difference when the tracking loop approaches loss of lock. In addition, considering the high short-term precision of the ODM solution and the need to preserve tracking-loop stability, it is recommended that the sensitivity gains and residual gate thresholds be tuned approximately inversely with $T_{\mathrm{ODM}}$ and $T$. Thus, when the ODM is updated more frequently, larger gains and gate thresholds can be used to grant greater authority to the ODM; when the update interval is longer, these parameters should be reduced accordingly. Establishing a theoretical basis and a practical tuning guide for given operating conditions is left for future work.

\section{Conclusion}
\label{sec:conclusion}

This paper presented an adaptive OAVT algorithm for orbit determination of LEO satellites under the practical CubeSat regime of single-frequency, code-based GNSS processing with broadcast ephemerides. The work was motivated by the gap between the robustness required by LEO-PNT applications and the limitations of existing spaceborne GNSS receiver architectures under degraded visibility, signal outages, corrupted measurements, and receiver off-duty intervals. To address this gap, a VTL-based receiver was combined with a low-fidelity ODM in a loosely coupled and computationally efficient framework, in which the ODM contributes position-velocity pseudo-measurements with adaptively assigned covariance based on GNSS stress and ODM--GNSS residual consistency. In this way, the proposed method preserves the frequent GNSS-based clock estimation and closed-loop tracking capability of VTL while providing physically informed orbital constraints when the GNSS solution becomes unreliable. The results show that, under nominal conditions, both VTL and OAVT substantially outperform STL, while OAVT remains essentially non-disruptive relative to standalone VTL. In particular, under nominal conditions, the OAVT 3D position RMSE is reduced from 2.631 m for STL to 1.081 m, corresponding to a 58.9\% improvement, while the 3D velocity RMSE is reduced from 0.3985 m/s to 0.1023 m/s, corresponding to a 74.3\% improvement. Under degraded conditions, however, the benefit of adaptive orbit aiding becomes decisive. OAVT maintains bounded and recoverable solutions during complete signal outages, gradual signal loss and reacquisition, and sustained low-visibility operation, whereas standalone VTL becomes increasingly vulnerable to cross-channel contamination and eventual loss of lock. Across the three degraded scenarios, the OAVT 3D position RMSEs are limited to 8.27, 10.17, and 26.65 m, respectively, while the corresponding 3D velocity RMSEs are limited to 0.53, 4.60, and 2.64 m/s. These gains are achieved with only moderate additional computational burden because OAVT preserves the dominant structure of the VTL receiver and introduces only a reduced-rate ODM branch. Overall, the proposed OAVT fills an important methodological gap for low-SWaP LEO onboard orbit determination by enabling more reliable state continuity and tracking resilience without sacrificing practical feasibility, thereby supporting CubeSat missions and future LEO-PNT systems that require timely and continuous satellite orbit information.

\appendix
\setcounter{table}{0}
\renewcommand{\thetable}{\thesection.\arabic{table}}
\section{ODM processing parameters}
\begin{table}[htbp]
\caption{ODM processing parameters in the current implementation \citep{mathlab2025odm}}
\label{tab:odm_parameters}
\centering
\begin{tabular}{|p{0.34\linewidth}|p{0.56\linewidth}|}
\hline
\textbf{Parameter} & \textbf{Setting} \\
\hline
Time reference & UTC Julian date (JD) \\
\hline
Propagator & Numerical (high precision)\\
\hline
Solver step size & Variable step (1 s - 86400 s) \\
\hline
Convergence tolerance & $10^{-8}$ \\
\hline
Gravitational model & Oblate ellipsoid (J2)\\
\hline
Atmospheric density model & NRLMSISE-00 \\
\hline
Third-body gravity & Sun \& Moon\\
\hline
Third-body ephemeris & DE432t \\
\hline
SRP shadow model & Dual cone \\
\hline
Spacecraft mass & 1.33 kg \\
\hline
Drag area & 0.01 m$^{2}$ \\
\hline
Drag coefficient $c_{d}$ & 1.05 \\
\hline
SRP area & 0.01 m$^{2}$ \\
\hline
Reflectivity coefficient $c_{R}$ & 1.21 \\
\hline
\end{tabular}
\end{table}

\section*{Acknowledgement}

The work described in this paper was fully supported by a grant from the Research Grants Council of the Hong Kong Special Administrative Region, China (PolyU15203326).

\section*{Data availability}
The data supporting the findings of this study are available from the corresponding author upon reasonable request.

\bibliographystyle{jasr-model5-names}
\biboptions{authoryear}
\bibliography{reference.bib}

\end{document}